\RequirePackage{fix-cm}
\documentclass[natbib]{svjour3}
\smartqed  
\journalname{}

\usepackage{setspace} 
\usepackage{amsmath,amssymb}
\usepackage{graphicx}
\usepackage{color}
\newcommand{\rot}[1]{{\color{black} #1}}
\newcommand{\blau}[1]{{\color{black} #1}}
\newcommand{\gruen}[1]{{\color{black} #1}}
\usepackage[normalem]{ulem}

\renewcommand{\cite}{\citep}

\newcommand{\be}{\begin{equation}} \newcommand{\ee}{\end{equation}} 
\newcommand{\ba}{\begin{eqnarray}} \newcommand{\ea}{\end{eqnarray}}

\newcommand{\vth}{V_\text{th}}
\newcommand{\vreset}{V_\text{r}}

\newcommand{\tl}{\hat{t}}
\newcommand{\tref}{t_\text{ref}}

\newcommand{\taum}{\tau_\text{m}}

\newcommand{\tauf}{\tau_\text{F}}
\newcommand{\taud}{\tau_\text{D}}

\newcommand{\var}[1]{\text{var}(#1)}
\newcommand{\Var}[1]{\text{Var}\left(#1\right)}

\newcommand{\lrrund}[1]{\!\left( #1 \right)}
\newcommand{\lreckig}[1]{\!\left[ #1 \right]}

\newcommand{\od}[2]{\frac{\mathrm{d}#1}{\mathrm{d}#2}}

\newcommand{\muxx}{Q}
\newcommand{\muuu}{P}
\newcommand{\muux}{R}

\begin{document}\sloppy

\title{Mesoscopic population equations for spiking neural networks with synaptic short-term plasticity
\thanks{This project received funding from the European Union's Horizon 2020 research and innovation programme under grant agreement No. 720270.}
%
}

\titlerunning{Mesoscopic population equations with synaptic short-term plasticity}        

\author{Valentin Schmutz         \and
        Wulfram Gerstner \and
        Tilo Schwalger
}


\institute{V. Schmutz \and W. Gerstner \and T. Schwalger\at
              Brain Mind Institute, \'{E}cole Polytechnique F\'{e}derale de Lausanne  (EPFL) Station 15, CH-1015 Lausanne, Switzerland\\
           \and T. Schwalger\at
\emph{Present address: Bernstein Center for Computational Neuroscience, 10115 Berlin, Germany\\
Institut f\"ur Mathematik, Technische Universit\"at Berlin, 10623 Berlin,Germany}\\
%
              \email{tilo.schwalger@bccn-berlin.de}           %
}

\date{Received: date / Accepted: date}

\maketitle
\begin{abstract}
Coarse-graining microscopic models of biological neural networks to obtain mesoscopic  models of neural activities is an essential step towards multi-scale models of the brain. Here, we extend a recent theory for mesoscopic population dynamics with {\em static} synapses to the case of {\em dynamic} synapses exhibiting short-term plasticity (STP). Under the assumption that spike arrivals at synapses have Poisson statistics, we derive analytically stochastic mean-field dynamics for the effective synaptic coupling between finite-size populations undergoing Tsodyks-Markram STP. The novel mean-field equations account for both finite number of synapses and correlations between the neurotransmitter release probability and the fraction of available synaptic resources. Comparisons with Monte Carlo simulations of the microscopic model show that in both feedforward and recurrent networks the mesoscopic mean-field model accurately reproduces stochastic realizations of the total synaptic input into a postsynaptic neuron \gruen{and accounts for stochastic switches between Up and Down states as well as for population spikes.} The extended mesoscopic population theory of spiking neural networks with STP may be useful for a systematic reduction of detailed biophysical models of cortical microcircuits to efficient and mathematically tractable mean-field models.     

\end{abstract}

\section{Introduction}

One of the primary goals in computational neuroscience is to
understand how higher brain functions arise from the interactions of
billions of nerve cells and their underlying biophysical processes at
the microscopic scale. Towards that goal, a crucial step is to develop
a theoretical framework that enables researchers to traverse models on
different levels of abstraction. An important example concerns the relation between simplified population rate models \cite{WilCow72} and biophysically detailed models at the microscopic scale. While mathematically tractable population rate models represent powerful frameworks for understanding a plethora of basic computations in cortical neural networks \cite[see e.g.,][]{BenBar95,WonWan06,BarTso07,ShpMor09,RubVan15}, they are heuristic models that lack a clear link to the underlying microscopic properties. On the other hand, highly detailed biophysical models of cortical microcircuits \cite{MarMul15} as well as simplified networks of point neurons \cite{IzhEde08,PotDie14,FieLan16} are closely linked to biophysical properties but lack mathematical tractability and do not provide a mechanistic understanding of emergent functional behavior. However, if we were able to systematically reduce biophysically detailed models to simplified networks of point neurons \cite{RosPoz16} and further to coarse-grained population rate models \cite{SchDeg17}, we \blau{might} be able to understand neural computations on the population level in terms of biophysical parameters. Combining \gruen{both steps in a bottom up approach} suggests a theory that links microscopic and \gruen{population rate models}.

A ubiquituous feature of cortical dynamics is synaptic short-term plasticity (STP) \cite{AbbVar97, MarWan98,DittKre00,ZucReg02}, i.e. dynamic changes of synaptic strength on time scales of 100~ms to 1000~ms induced by presynaptic neural activity. Theoretical studies have shown that STP exerts profound effects on network activity \cite{LevHer07,PitIba17} and information processing capabilities \cite{AbbVar97,ForRos01,MerLin10,RosRub12,DroSch13}. In particular, in a recent biophysically detailed microcircuit model \cite{MarMul15}, STP has been a critical factor for reproducing experimentally observed activity patterns. Therefore, a faithful reduction to population rate models should incorporate the effect of STP. Mean-field descriptions for populations of dynamic synapses are central for such a reduction. Present mean-field theories for STP have been developed for the case of infinitely large (``macroscopic'') populations \cite{TsoPaw98,HolTso06,BarTso07,MonBar08}. In mesoscopic neural circuits such as cortical columns, the number of cells are often only on the order of a few hundred or thousand cells per neuronal population \cite{LefTom09} necessitating a finite-size description. A systematic finite-size mean-field theory for dynamic synapses is, however, currently lacking.

Recently, a mesoscopic mean-field model \gruen{derived} from a microscopic model of interacting spiking neurons with {\em static} synapses \gruen{has been proposed} \cite{SchDeg17}. Here, we build upon this work and derive an extension of the mesoscopic model that includes the effect of {\em dynamic} synapses due to STP at the microscopic level. To this end, we first study in Sec.~\ref{sec:poisson} a feedforward setup \cite{LinGan09,MerLin10}, in which a postsynaptic neuron is driven by a presynaptic population of $N$ Poisson neurons each connected to the postsynaptic neuron via a dynamic synapse. We derive a set of equations for an effective ``mean-field synapse'' that approximates the effect of the $N$ synapses on the total synaptic input current \gruen{in terms of} the first- and second-order statistics of the total postsynaptic input. In Sec.~\ref{sec:recurr-netw-with}, we first illustrate how the mesoscopic STP model accurately replicates population spikes and switches between Up and Down states  exhibited by a recurrent  network of Poisson neurons. We then incorporate the mesoscopic STP model into our previous mesoscopic population model \cite{SchDeg17} for generalized integrate and fire (GIF) neurons. We show that the resulting extension faithfully reproduces population spikes observed in a microscopic simulation despite the fact that GIF neurons have non-Poisson statistics. Finally, in Sec.~\ref{sec:discuss}, we discuss the limitations of our mesoscopic model for GIF neurons and mention possible  theoretical extensions. Detailed derivations of the mean-field equations are presented in the Appendix.

\section{Feedforward network with a finite number of dynamic synapses}
\label{sec:poisson}

\begin{figure*}
  \centering
  \includegraphics[width=1.\textwidth]{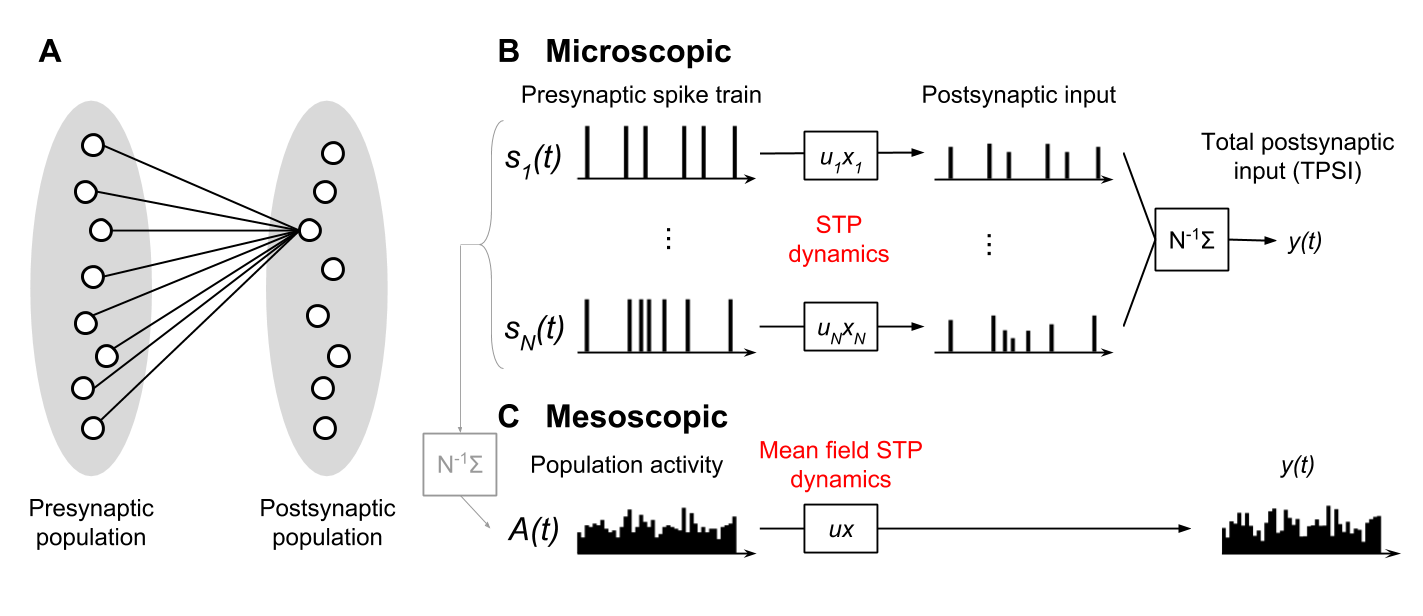}
  \caption{Illustration of the setup in the feedforward network. (A) Two populations connected in a feedforward manner via dynamic synapses. We focus on the connections from neurons \gruen{$j$, $j=1,\dotsc,N$}, in the presynaptic population to a specific postsynaptic neuron \gruen{$i$}. (B) Microscopic picture of $N$ presynaptic spike trains $s_j(t)$ driving the STP dynamics of $u_j(t)$ and $x_j(t)$ for each of the $N$ synapses. The postsynaptic input resulting from synapse $j$ is $u_j(t)x_j(t)s_j(t)$ and the total postsynaptic input is $y(t)=N^{-1}\sum_{j=1}^Nu_j(t)x_j(t)s_j(t)$. (C) Mesoscopic picture of {\em one} effective synapse with {\em mean field} STP dynamics driven by the population activity $A(t)$ of $N$ neurons. The population activity $A(t)$ is defined as the population average of the spike trains of each of the $N$ neurons forming the population. Thus, when the individual spike trains $s_j(t)$ are known, $A(t)=N^{-1}\sum_{j=1}^N s_j(t)$.}
  \label{fig:illu1}
\end{figure*}

\gruen{Feedforward pathways exhibiting STP are prominent in the nervous system.} Examples include visual \cite{AbbVar97}, auditory \cite{CooSch03}, somatosensory \cite{HigCon06} and periform \cite{OswUrb12} cortices. Thus, we first study a feedforward network:  $N$ neurons from
a presynaptic population are connected to \gruen{a given postsynaptic neuron via $N$ synapses} (Fig.~\ref{fig:illu1}A). Our aim is to reduce the
$N$ synaptic couplings to a single synapse that effectively connects
the presynaptic population as a whole to the postsynaptic neuron. In
other words, the microscopic model of $N$ \gruen{synapses driven by $N$ presynaptic spike trains} will be approximated by a \gruen{single set
of mesoscopic equations for the total postsynaptic input}. To simplify the mathematical analysis, we assume here that
the presynaptic spiking is a Poisson process.

\subsection{Microscopic model}

We are interested in the synaptic current $I_{syn}(t)$ (or a synaptic conductance $g_{syn}(t)$) into the postsynaptic neuron. Synaptic inputs are filtered by the conductance dynamics of the synapse, which \gruen{we model} as 
\begin{equation}
\label{eq:1}
   I_\text{syn}(t)=J\int_{-\infty}^t\epsilon(t-t')y(t')\,dt',
\end{equation}
where $J$ is the effective synaptic weight (in units of electrical charge) \gruen{which is identical for all synapses onto the postsynaptic neuron}, $\epsilon(t)$ is a synaptic filtering kernel (defined as the postsynaptic current normalized by the total charge elicited by a single spike) and $y(t)$ is the {\em total postsynaptic input} (TPSI) before synaptic filtering.  For synapses with STP, the TPSI can be modeled by
\begin{equation}
  y(t)=\frac{1}{N} \sum_{j=1}^N R_j(t)s_j(t),
\label{eq:post-input}
\end{equation}
where $s_j(t)=\sum_{t_k^j}\delta(t-t_k^j)$, $j=1,\dotsc,N$, denotes
the Dirac-delta spike train of neuron $j$ in the presynaptic population. \gruen{We assume that the} firing times $\{t_k^j\}_{k\in\mathbb{Z}}$ of neuron $j$ occur
independently according to a Poisson process with rate $r(t)$. The
normalization factor $1/N$ in Eq.~(\ref{eq:post-input}) ensures that $y$ is an intensive mesoscopic quantity such that its mean is independent of population size. In Eq.~(\ref{eq:post-input}),
the amplitudes of the spikes are modulated by a factor $R_j$, which mediates the effect of STP. The variable $R_j(t)$ represents the fraction of neuro-transmitters that are released upon a spike at time $t$ \cite{TsoPaw98}. Thus, $y(t)$ can be interpreted as the population-averaged amount of neurotransmitter released at time $t$; it is the essential quantity that drives the dynamics of the postsynaptic current. To model the modulation $R(t)$, we choose the Tsodyks-Markram model
\cite{TsoPaw98,MonBar08}, which is governed by two variables, the
amount of available resources $x_j$ and the utilization (or release probability)
$u_j$, for each synapse $j$. Their product defines the modulation factor:
\begin{equation}
  R_j=u_j x_j,
\end{equation}
Given the presynaptic spike trains
$s_j(t)$, these variables obey the dynamics
\begin{align}
  \label{eq:stpmicro_u}
  \od{u_j}{t}&=\frac{U_0-u_j}{\tau_F}+U(1-u_j)s_j(t),\\
  \od{x_j}{t}&=\frac{1-x_j}{\tau_D}-u_jx_js_j(t),
  \label{eq:stpmicro_x}
\end{align}
where $\tauf$ and $\taud$ are the facilitation and depression time
constants, respectively, $U_0$ is the baseline utilization of synaptic
resources, $U$ determines the increase in the utilization of synaptic
resources by a spike. Here, all variables multiplying $s_j(t)$ have
to be evaluated at time $t^-$, i.e. immediately before the delta
spikes. 

\subsection{Theory for mesoscopic synaptic dynamics}
\label{sec:meso-var}

We want to find a mesoscopic dynamics that determines $y(t)$, and hence the synaptic current $I_{syn}(t)$. This means that we look for a system of evolution equations that should not explicitly depend on microscopic variables such as individual spike trains $s_j$ or synaptic variables $u_j$ and $x_j$. A central quantity that has been used in mesoscopic population equations \cite{SchDeg17} is the population activity defined as the superposition of spike trains:
\begin{equation}
  \label{eq:pop-ac}
  A(t) = \frac{1}{N}\sum_{j=1}^Ns_j(t).
\end{equation}
In our case, $A(t)$ is the population activity of the presynaptic population of Poisson processes $\{s_j(t)\}$. Thus, we expect that the population activity will play the role of an external drive to the mesoscopic synaptic dynamics. 

To relate the TPSI, Eq.~\eqref{eq:post-input}, to the presynaptic population activity we consider  the collection $\{\hat{t}_k\}_{k\in\mathbb{Z}}$  of all spike times of the superposition spike train $\sum_j s_j$. In terms of the spike times $\hat{t}_k$ the population activity and the TPSI can be written as $A(t)=\frac{1}{N}\sum_k\delta(t-\hat{t}_k)$ and
\begin{equation}
  \label{eq:yn-1}
  y(t)=\frac{1}{N}\sum_{k}u_{j(k)}(t)x_{j(k)}(t)\delta(t-\hat{t}_k),
\end{equation}
respectively. In Eq.~\eqref{eq:yn-1}, the index $j(k)$ points to the synapse that receives a spike at time $\hat{t}_k$ (note that the probability that two synapses receive a spike at exactly the same time is zero). Equation \eqref{eq:yn-1} still contains the microscopic variables $u_j$ and $x_j$. To remove this dependence we note that the Poissonian spike times $\hat{t}_k$ \rot{independently} sample different synapses $j(k)$ in a uniform manner. In other words, given an input spike at time $\hat{t}_k$, the likelihood that this spike arrives from presynaptic neuron $j$ is \gruen{identical} for all presynaptic neurons $j=1,\dotsc,N$ because the instantaneous firing probability is the same for all Poisson processes\footnote{This is no longer true for non-Poissonian spike trains. To see this, consider the extreme case that neurons fire regularly (but asynchronously) with period $T$. The arrival of a spike at time $t$ means that the respective synapse has seen its previous spike at time $t-T$. This causes a bias in the variables $u$ and $x$ at that synapse because these variables decayed towards baseline during a maximal interval $(t-T,t)$ after their last jump. Specifically, $u$ ($x$) will be smaller (larger) than the respective population average. \rot{Hence, synapse identies at spike events will depend on synaptic variables $u$ and $x$ and can therefore not be regarded as independent and uniformly distributed samples from the set of indices $\{j\in\mathbb{N}|1\le j\le N\}$.}}. Thus, upon spiking the corresponding synaptic variables $u_j$ and $x_j$ are samples from the current joint distribution $p(u,x,t)=\frac{1}{N}\sum_j\delta(u-u_j(t))\delta(x-x_j(t))$ of synaptic variables across the population. \gruen{Note that for a given synapse $j$, the variables $u_j$ and $x_j$ are correlated because they are both driven by the same spike train $s_j(t)$.} From the mesoscopic point of view, \rot{single spike trains $s_j(t)$ are not known, only the summed activity $A(t)$ is available. Thus,} a specific \rot{realization} $j(k)$ \rot{associating} synapse identities to spike times is irrelevant as long as \rot{$(u_{j(k)},x_{j(k)})$ is a sample} from $p(u,x,\hat{t}_k)$. Therefore, the variables $u_{j(k)}$ and $x_{j(k)}$ can be replaced by correlated random variable $\hat{u}(\hat{t}_k)$ and $\hat{x}(\hat{t}_k)$ \rot{that are jointly} sampled from $p(u,x,\hat{t}_k)$. With these variables, Eq.~\eqref{eq:yn-1} can be rewritten as
\begin{align}
  y(t)&=\frac{1}{N}\sum_{k}\hat{u}(\hat{t}_k)\hat{x}(\hat{t}_k)\delta(t-\hat{t}_k),\nonumber\\
&=\hat{u}(t)\hat{x}(t)A(t).  \label{eq:yn-approx}
\end{align}
\rot{
Note that the variables $\hat{u}(t)$ and $\hat{x}(t)$ are only evaluated at spike times $\{\hat{t}_k\}$.

It remains to obtain a closed statistical description for the stochastic variables $\hat{u}(t)$ and $\hat{x}(t)$. To this end, we employ a moment-closure approximation, in which joint cumulants of order three and higher are discarded. This Gaussian approximation requires the first and second-order moments of $\hat{u}(t)$ and $\hat{x}(t)$ conditioned on the current microstate $\{u_j(t),x_j(t)\}_{j=1}^N$, which can be well estimated using the population average:
\begin{subequations}
  \label{eq:mus}
\begin{align}
  \label{eq:muu}
  u(t)&:=\frac{1}{N}\sum_{j=1}^Nu_j(t),\\
  \label{eq:mux}
  x(t)&:=\frac{1}{N}\sum_{j=1}^Nx_j(t),\\
  \label{eq:muuu}
  \muuu(t)&:=\frac{1}{N}\sum_{j=1}^Nu_j^2(t),\\
  \label{eq:muxx}
  \muxx(t)&:=\frac{1}{N}\sum_{j=1}^Nx_j^2(t),\\
  \label{eq:muux}
  \muux(t)&:=\frac{1}{N}\sum_{j=1}^Nu_j(t)x_j(t).
\end{align}
\end{subequations}
Importantly, these quantities serve as new mesoscopic variables in our second-order mean-field theory of STP. They allow us to model  $\hat{u}(t)$ and $\hat{x}(t)$ as sums of their means and Gaussian fluctuations: }
\begin{subequations}
  \label{eq:gaussian}
  \begin{align}
    \label{eq:gaussian-u}
    \hat{u}(t) &= u(t)+\varepsilon_u(t),\\
    \label{eq:gaussian-x}
    \hat{x}(t) &= x(t)+\varepsilon_x(t).
  \end{align}
\end{subequations}
\rot{Here, the components of the vector} $\varepsilon(t)=[\varepsilon_u(t),\varepsilon_x(t)]^T$ are \gruen{zero-mean} Gaussian noise  with covariance matrix
\begin{gather}
  \label{eq:cova}
\langle\varepsilon(t)\varepsilon^T(t')\rangle=\Sigma(t)
\delta_{t,t'},\nonumber\\
\Sigma(t)=  \begin{pmatrix}
    \muuu(t)-u^2(t)&\muux(t)-u(t)x(t)\\
    \muux(t)-u(t)x(t)&\muxx(t)-x^2(t)
  \end{pmatrix}
\end{gather}
where $\delta_{t,t'}$ is the Kronecker delta, which is unity if $t=t'$ and zero otherwise.

\gruen{The dynamics of the mesoscopic variables $u$, $x$, $P$, $Q$ and $R$ are given by first-order differential equations:} 
\begin{subequations}
  \label{eq:meso-full}
\begin{align}
  \label{eq:meso-u}
  \od{u}{t}&=\frac{U_0-u}{\tau_F}+U(1-\hat{u})A(t),\\
  \label{eq:meso-x}
  \od{x}{t}&=\frac{1-x}{\tau_D}-\hat{u}\hat{x}A(t)
\end{align}
for the means and
\begin{align}
  \label{eq:meso-P}
  \od{\muuu}{t}&=2\frac{U_0u-\muuu}{\tau_F}+U(1-\hat{u})[(2-U)\hat{u}+U]A(t),\\
  \label{eq:meso-Q}
  \od{\muxx}{t}&=2\frac{x-\muxx}{\tau_D}-\hat{u}\hat{x}^2(2-\hat{u})A(t),\\
  \label{eq:meso-R}
  \od{\muux}{t}&=\frac{U_0x-\muux}{\tau_F}+\frac{u-\muux}{\tau_D}+\hat{x}\left[U(1-\hat{u})^2-\hat{u}^2\right]A(t)
\end{align}
\end{subequations}
for the second-order \rot{moments}. Equations \eqref{eq:yn-approx}, \eqref{eq:gaussian} -- \eqref{eq:meso-full} completely determine the mesoscopic dynamics of the TPSI and constitute the main result of the paper. \gruen{The complete derivation is in the Appendix Sec.~\ref{sec:meso-eq}}. The derivation of Eqs.~\eqref{eq:meso-full} \rot{corresponds to} a second-order moment-closure approximation. If we do a first-order moment-closure, the variance and covariance of $\hat{u}(t)$ and $\hat{x}(t)$ are neglected ({\em i.e.} $\epsilon(t) = 0$) and we obtain simply
\begin{subequations}
  \label{eq:first-full}
\begin{align}
  \od{u}{t}&=\frac{U_0-u}{\tau_F}+U(1-u)A(t),\\
  \od{x}{t}&=\frac{1-x}{\tau_D}- u x A(t).
\end{align}
\end{subequations}
These are the classic mean-field equations derived for $N\rightarrow\infty$ by \citet{TsoPaw98}.
In this paper, we name Eqs.~\eqref{eq:meso-full} the second-order mean field theory (abbreviated 2nd-order MF) and Eqs.~\eqref{eq:first-full} the first-order mean-field theory (abbreviated 1st-order MF). \gruen{In the following, we call $R$ the modulation factor since it characterizes the mean of the product $\hat{u}\hat{x}$ which modulates synaptic efficacies.}

\begin{figure*}[t]
    \centering
    \includegraphics[trim={4.5cm 0 0 0}, scale=.275]{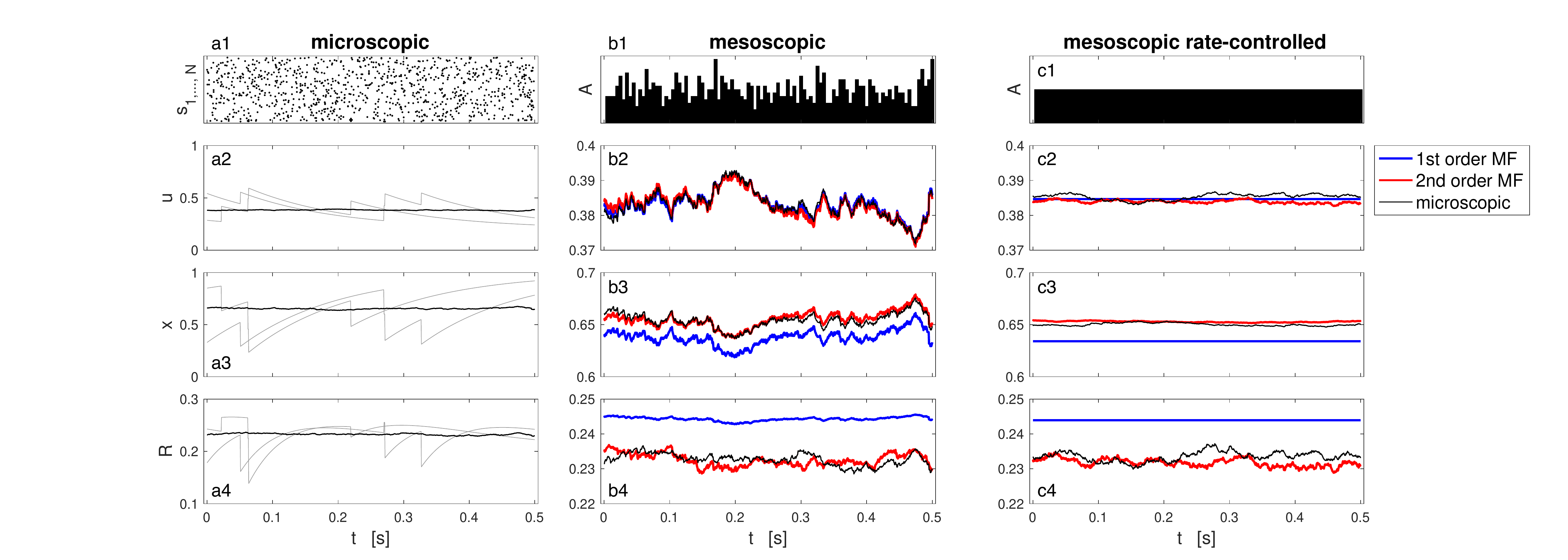}
    \caption{\textbf{Example of microscopic and mesoscopic synaptic dynamics for 200 presynaptic stationary Poisson neurons.} (a1) Raster plot of $N = 200$ presynaptic stationary Poisson neurons with rate 10~Hz. (a2-4) Trajectories of variables $u_j(t)$ and $x_j(t)$ and the resulting release probability $R_j(t) \equiv u_j(t)x_j(t)$ for two example neurons (gray lines). The black line shows the population averages $u(t)$, $x(t)$ and $R(t)$ calculated from Eqs.~\eqref{eq:mus}. (b1) Population activity $A(t)$ corresponding to the 200 spike trains shown in (a1). (a2-a4) Trajectories of the mesoscopic variables $u(t)$, $x(t)$ and $R(t)$ predicted by the 1st and 2nd-order MF (blue and red, respectively) compared to the microscopic simultation (black) which correspond to the population averages shown on the left. Note that the y-axis scale is different in (a2-4) and (b2-4). In (b4), we see that, while finite-size fluctuations in $R$ for the population average are reproduced by both 1st and 2nd-order MF, the 1st-order MF makes an error in predicting the mean. \gruen{(c1-4) is the same as (b1-4) except that we force $A$ to be constant: while the $s_j$ are still Poisson spike trains with rate 10 Hz, they are generated such that $A$ is constant over time. This removes the effect of the finite-size fluctuations of $A$ on the finite-size fluctuations of the mesoscopic STP $u$, $x$ and $R$. (b2-b4) In contrast with the 1st-order MF, the 2nd-order MF reproduces the residual finite-size fluctuations observed in the microscopic simulation}. Synaptic parameters : $\taud = 0.15$ s, $\tauf = 0.15$ s, $U = U_0 = 0.2$. \gruen{(b1 and c1) are binned with bin size 0.005 ms.}}
    \label{fig:example}
\end{figure*}

In the appendix Sec.~\ref{sec:num_imp}, we provide an efficient simulation algorithm for Eqs.~\eqref{eq:meso-full}. Trajectories of $u_j$, $x_j$ and $R_j$ as well as $u$, $x$ and $R$ obtained from a microscopic simulation are shown in Fig.~\ref{fig:example}. The mesoscopic variable $R$ is tracked by the 2nd-order MF dynamics with high accuracy (Fig.~\ref{fig:example}b4). The 1st-order MF also yields reasonable results, although \gruen{a small deviation in the mean $R$ over time} is apparent (Fig.~\ref{fig:example}b4). This is consistent with previous findings \cite{TsoPaw98}, where it has been shown analytically that in the stationary case relative correlations are small, \gruen{but significant}. \gruen{Note that the 2nd-order MF distinguishes two sources of finite-size noise: noise that comes from the finite-size fluctuations of $A$ and second, noise that comes from the sampling of $\hat{u}$ and $\hat{x}$ at each spike. This second source of noise is absent in the 1st-order MF. In a numerical simulation with time step $\Delta t$, it is possible to isolate this second source of noise: if $N \cdot \Delta t \cdot r$ (where $r$ is the rate of the Poisson process) is a strictly positive integer $\alpha$, we can choose, independently at each time step, $\alpha$ neurons uniformly across the $N$ neurons and make them spike. This procedure generates $N$ discretised Poisson spike trains of rate $r$ with a constant population activity $A$ over time. Note that in this case, the spike trains are not independent of each other but this does not affect our derivation. This procedure is followed in Fig.~\ref{fig:example}c1-4: the 1st MF predicts noseless STP dynamics whereas the 2nd-order MF accurately reproduces the residual finite-size fluctuations.}

\begin{figure}[t]
  \centering
  \includegraphics[scale=0.55]{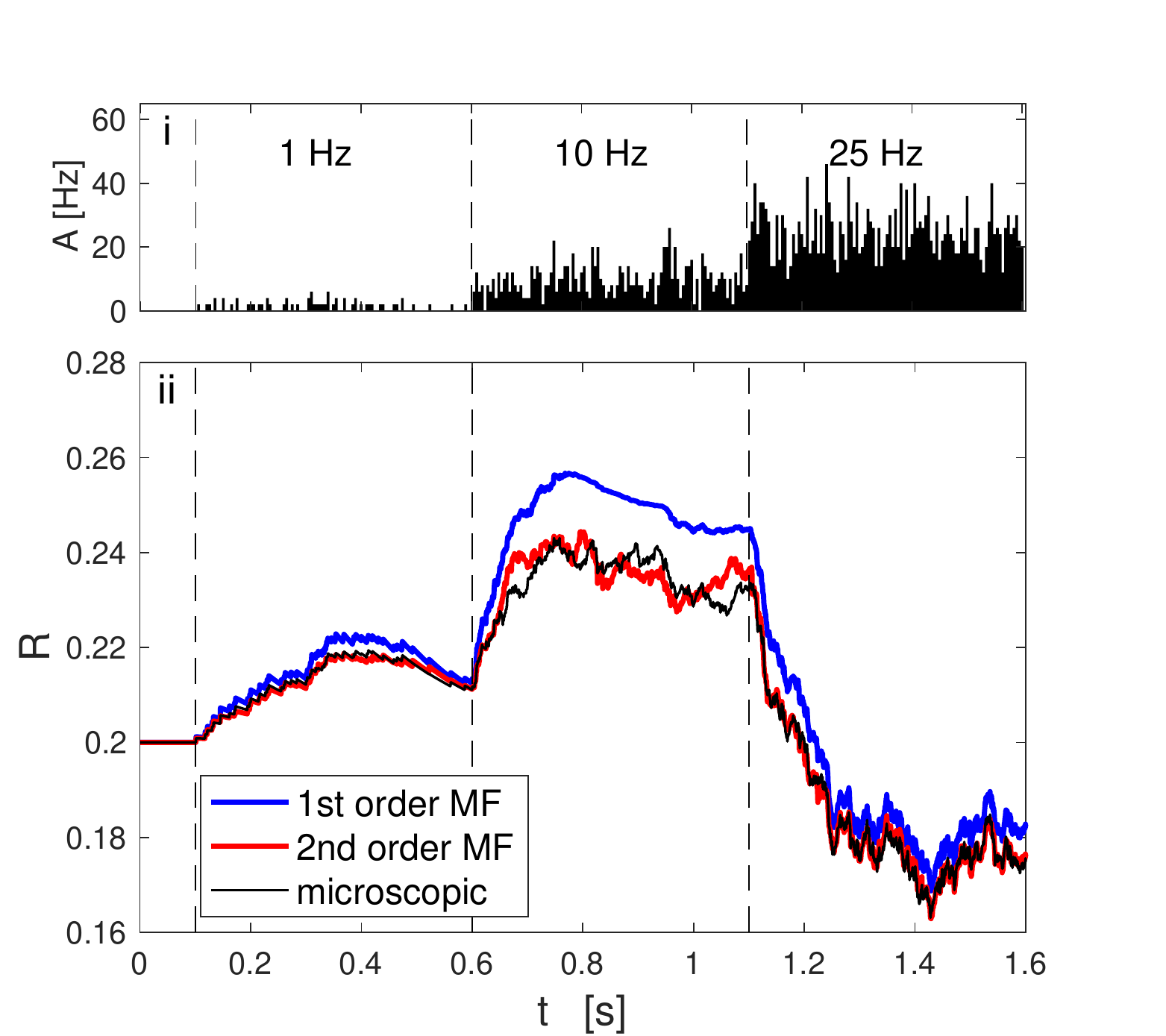}
  \caption{\textbf{Synaptic dynamics in response to step increments in the the presynaptic firing rate.} \gruen{(i) Population activity $A$ of 100 Poisson neurons when the presynaptic firing rate changes sharply from 0 Hz to 1, 10 and 25 Hz. (ii) Corresponding mesoscopic release probability $R$ predicted by the 1st and 2nd-order MF (blue and red lines respectively) compared to the microscopic simulation (black line).} Note that at 10 Hz the 2nd-order MF corrects the overestimation in the mean of the 1st-order MF and reproduces finite-size fluctuations of amplitude similar to that of the population average. Synaptic parameters: $\tau_D=0.15$ s, $\tau_F=0.15$ s, $U=U_0=0.2$. }
  \label{fig:step-resp}
\end{figure}

The deviations of the 1st-order MF become more pronounced during non-stationary transients \gruen{caused by stepwise increases of the rate of the Poisson process} (Fig.~\ref{fig:step-resp}). The response to step increases accurately traced by the 2nd-order MF, \gruen{but not by the 1st-order MF which neglects the correlations between $u_j$ and $x_j$.} 

\subsection{Statistics of the total postsynaptic input}
\label{sec:tpsi-stat}

\begin{figure*}[t]
    \centering
    \includegraphics[width=0.9\textwidth]{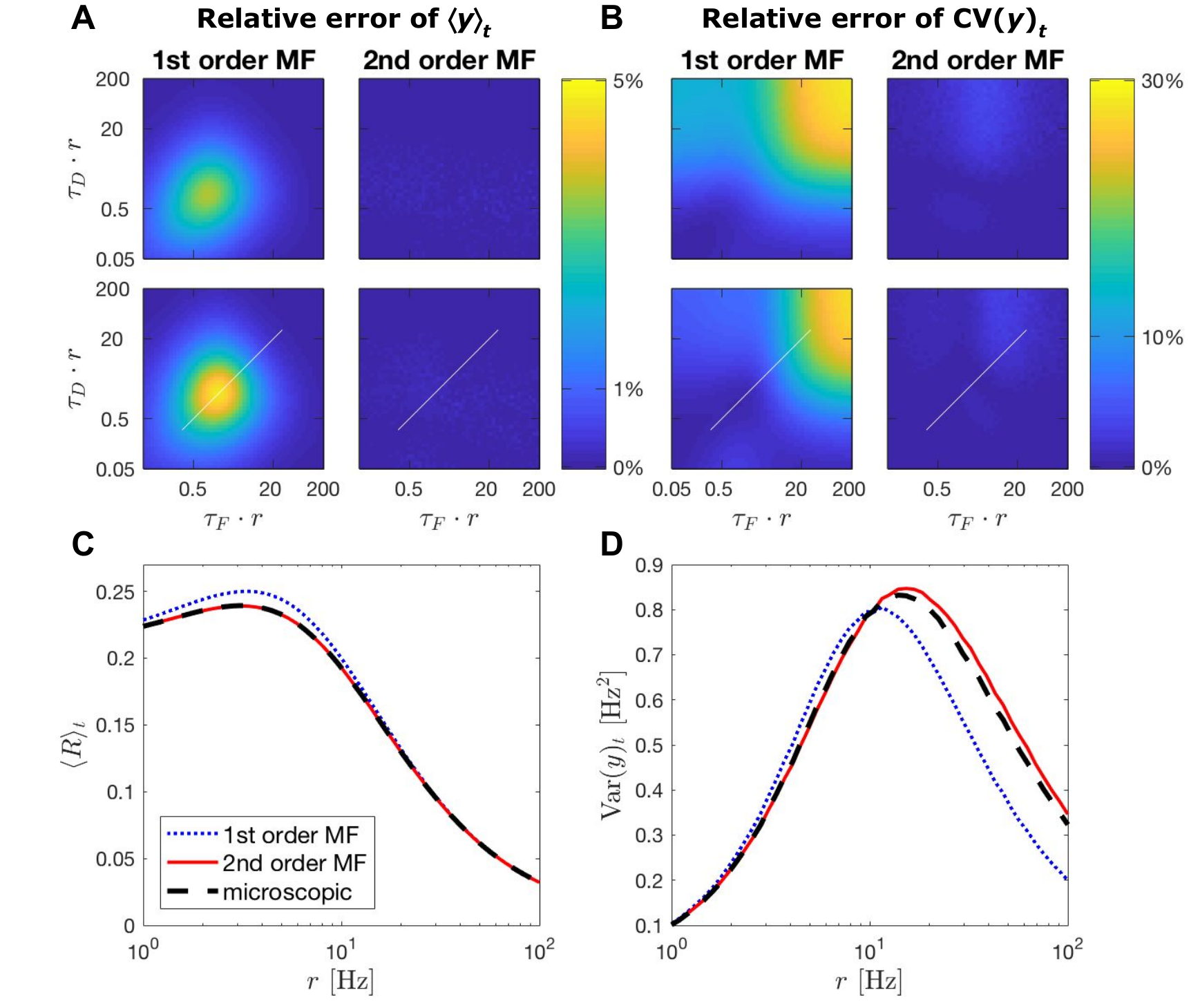}
    \caption{\textbf{First- and second-order statistics of the TPSI ($y$) for a presynaptic population of 100 stationary Poisson neurons.} (A, B) Relative error of the mean TPSI over time $\langle y \rangle_t$ (A) and the coefficient of variation of the TPSI over time ($\text{CV}(y)_t$) (B) predicted by the 1st and 2nd-order MF (left and right column respectively) with respect to microscopic simulation, as a function of the synaptic parameters $\taud$, $\tauf$ and \gruen{for two values of $U$ (with $U=U_0$);} $U$ is set to 0.5 on the upper row and 0.2 on the lower row. On the x- and y-axes, $\tauf \cdot r$ is a unitless quantity. In (A), the maximum relative error is 4.7\% for the 1st-order MF and 0.3\% for the 2nd-order MF. In (B), the maximum relative error is 28.6\% for the 1st-order MF and 4.0\% for the 2nd-order MF. As scaling $\tauf$ and $\taud$ is equivalent to scaling the firing rate $r$, the relative error at different firing rates can be read moving along the diagonal \gruen{(white line)}. (C) Mean modulating factor $\langle R \rangle_t$ over time predicted by the 1st and 2nd-order MF (dotted blue and solid red lines respectively) compared to microscopic simulations (dashed black line) as a function of the firing rate $r$ for a specific set of synaptic parameters. (D) TPSI variance over time ($\Var{y}_t$) predicted by the 1st and 2nd-order MF compared to microscopic simulations as a function of the firing rate $r$ for a specific set of synaptic parameters. Synaptic parameters used in (C-D) correspond to the white line in (A-B) and are: $\taud = 0.3$ s, $\tauf = 0.3$ s and $U =U_0=0.2$. Simulation time step is 0.5 ms.}
    \label{fig:params}
\end{figure*}

\begin{figure*}[t]
  \centering
  \includegraphics[width=\linewidth]{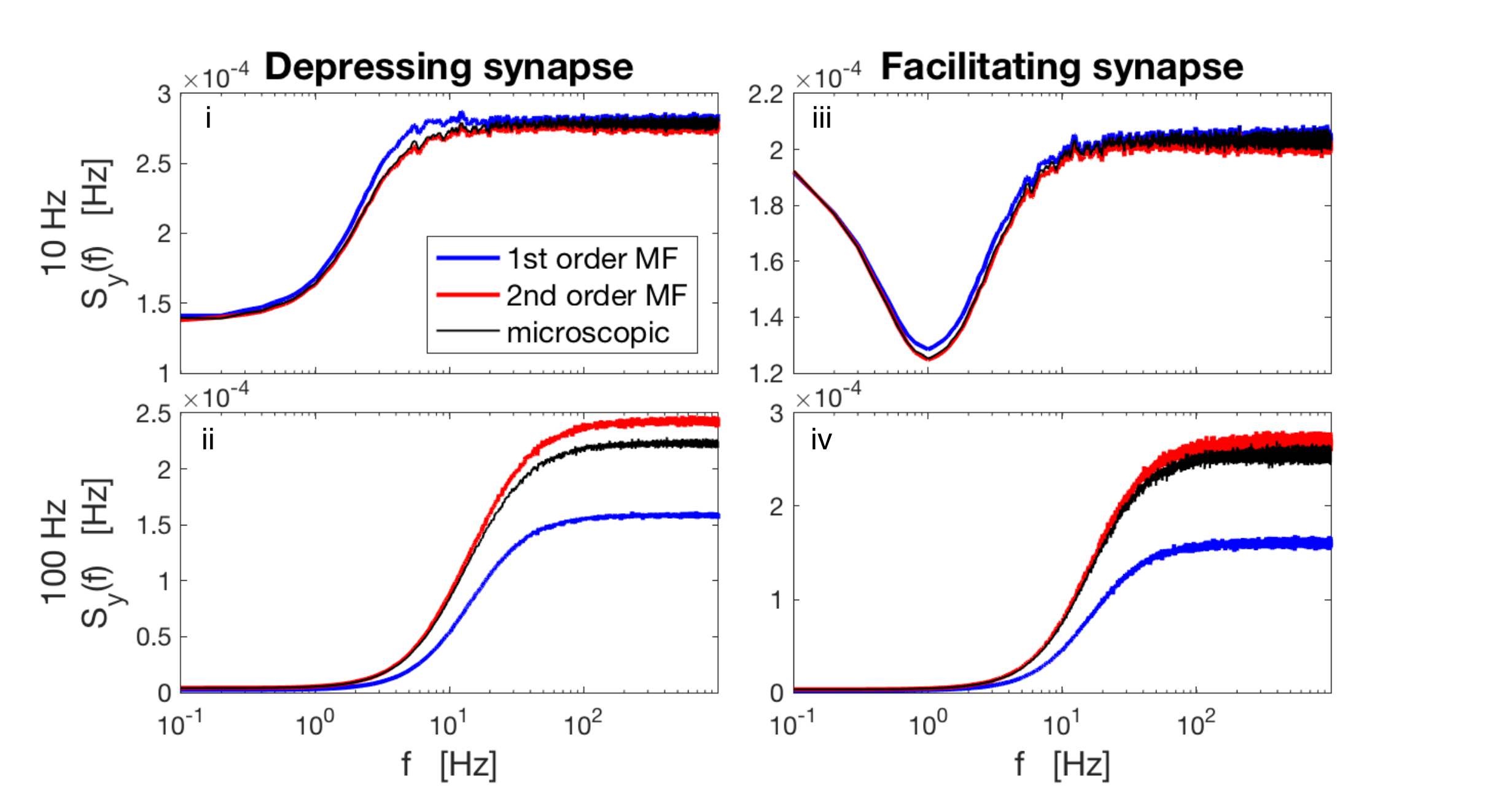}
  \caption{\textbf{Power spectral densities (PSD \gruen{of the TPSI} given a presynaptic population of 5000 stationary Poisson neurons.} PSD of the TPSI of a depressing synapse (i and ii) and a facilitating synapse (iii and iv), when the firing rate \gruen{of the presynaptic neurons} is 10~Hz (i and iii) and 100 Hz (ii and iv), predicted by the 1st and 2nd-order MF (blue and red lines respectively) compared microscopic simulations (black line). Each PSD is averaged over 5000 simulations and further smoothed using a moving average. Parameters for the depressing synapse: $\tau_D=0.1$ s, $\tau_F=0.05$ s, $U=U_0=0.5$. Parameters for the facilitating synapse: $\tau_D=0.1$ s, $\tau_F=0.7$ s, $U=U_0=0.1$. \gruen{$A$ in the upper panel is binned with bin size 0.005 ms.}}
    \label{fig:psd-feedforward}
\end{figure*}

To compare microscopic and mesoscopic descriptions more systematically, we measured the first- and second-order  statistics from simulations for varying parameters. At first, we computed the mean of the modulation factor  $R$ for the stationary process \gruen{($\langle s_j(t) \rangle = r =\text{const.}$) using the microscopic dynamics,}
\begin{equation}
  \label{eq:mean-tpsi}
  \langle R\rangle=\langle u_j(t)x_j(t)\rangle,
\end{equation}
where $\langle\cdot\rangle$ denotes the ensemble (trial) average, i.e. the average over realizations of the Poisson processes $s_j(t)$. The mean TPSI  is \gruen{proportional} to the mean modulation factor \gruen{because} $\langle y\rangle=\frac{1}{N}\sum_j\langle u_jx_j\rangle\langle s_j\rangle=\langle R\rangle r$. This simple proportionality follows from the fact that $u_j(t)x_j(t)$ \gruen{at time $t$} is
uncorrelated with $s_j(t)$ at \gruen{at the same time} because of the Poisson
statistic \gruen{of the spike train and the update of variables after a spike; cf. Eqs.~\eqref{eq:stpmicro_u} and \eqref{eq:stpmicro_x}}. As known from previous work \cite{TsoPaw98}, the mean modulation $\langle R \rangle$ of a facilitating synapse increases with increasing firing rate of presynaptic neurons when firing rates are small, and  decreases again at high rates due to depression (Fig.~\ref{fig:params}C).
The 1st-order MF shows small deviations in the mean modulation $\langle R\rangle$, which are \gruen{removed} by the 2nd-order MF (Fig.~\ref{fig:params}C). A closer inspection of the full parameter regime reveals that the deviation of 1st-order MF never exceed $5\%$ (Fig.~\ref{fig:params}A). Therefore the stationary mean TPSI is sufficiently well explained by the 1st-order MF.
 
\gruen{Also} we compared the statistics of fluctuations of the TPSI by measuring the respective  power spectral densities (PSD). The PSD can be computed as  
\begin{equation}
  S_{yy}(f)=\frac{\langle \tilde{y}^*(f)\tilde{y}(f)\rangle}{T},
\end{equation}
where $\tilde{y}(f)=\int_0^Tdt\,\exp(2\pi ift)y(t)$ denotes the Fourier transform of $y(t)$ for a finite \gruen{but large enough} time window $T$. We found that the 2nd-order MF significantly better captured the variance (Fig.~\ref{fig:params}D) and the PSD (Fig.~\ref{fig:psd-feedforward}) of the stationary fluctuations than the 1st-order MF. A closer inspection of the coefficient of variation of the fluctuations, $\sqrt{\Var{y}}/\langle y\rangle$, over the full parameter space revealed that the 1st-order MF deviated up to 30\% (especially for slow synaptic dynamics or high rates, Fig.~\ref{fig:psd-feedforward}(ii,iv)), whereas the 2nd-order model performed well in the whole parameter space (Fig.~\ref{fig:params}B). We should specify that the error of the 1st-order MF is negative, \textit{i.e.} the 1st-order MF underestimates the coefficient of variation up to 30\%. This comes from the fact that in the mesoscopic equations for the 1st MF Eqs.~\eqref{eq:first-full}, finite-size fluctuations of the mesoscopic variables $u$ and $x$ are ignored.

In conclusion, \gruen{while} mean responses for stationary cases are well captured by the 1st-order MF, \gruen{the} 2nd-order MF \gruen{gives a} significantly better description of transient responses (Fig.~\ref{fig:step-resp}) and fluctuations (Figs.~\ref{fig:params}B,D and \ref{fig:psd-feedforward}).

\section{Recurrent network with STP}
\label{sec:recurr-netw-with}


\subsection{Microscopic model}
\label{sec:micro-rec}

As shown in our previous work \cite{SchDeg17}, networks of multiple interacting homogeneous populations of spiking neurons can be accurately predicted with a mesoscopic model. In order to assess whether our mesoscopic theory of STP could be incorporated in this general model, we consider for simplicity the special case of a single population, the network architecture \gruen{taken as} random with fixed in-degree $C=pN$, where $N$ is the number of neurons in the population and $p$ is the connection probability. The synaptic strength is constant with magnitude $w$ (in mV). The TPSI $y_i(t)$ and the synaptic current $I_{\text{syn},i}(t)$ driving the postsynaptic neuron $i$ and are related by Eq.~\eqref{eq:1} with $J=\frac{\taum}{R_\text{m}}Npw$. In this paper, we use a synaptic filtering kernel with instantaneous rise and exponential decay  corresponding to the first-order kinetics
\begin{equation}
	\label{eq:I_syn}
    \tau_s \od{I_{\text{syn},i}}{t} = -I_{\text{syn},i}+J y_i(t), 
\end{equation}
where $\tau_s$ is the synaptic filtering time constant. Importantly, the effect of STP is contained in the TPSI $y_i(t)=\frac{1}{C}\sum_{j\in\Gamma_i}u_j(t)x_j(t)s_j(t)$ via the synaptic variables $u_j$ and $x_j$ given by the dynamics Eq.~\eqref{eq:stpmicro_u} and \eqref{eq:stpmicro_x}. Here, $\Gamma_i$ denotes the index set of presynaptic neurons that connect to neuron $i$.

As our derivation of the mesoscopic theory of STP uses the assumption that neurons have Poisson statistics, we first apply our theory to Poisson rate neurons, which do not exhibit a dependence on spike history. Then, using the same setup, the theory will be applied to a network of generalized integrate-and-fire (GIF) neurons with pronounced spike-history effects.

\subsubsection{Poisson rate model}
The Poisson rate model, is defined by a continuous variable $h_i(t)$, called input potential. The input potential \gruen{of neuron $i$} obeys the dynamics
\begin{equation}
  \label{eq:rate-model}
  \taum\od{h_i}{t}=-h_i+\mu(t)+ R_\text{m}I_{\text{syn},i}(t),
\end{equation}
where $\taum$ represents the membrane time constant and
$\mu(t)=V_{\text{rest}}+R_\text{m} I_{\text{ext}}(t)$ is the
total drive in the absence of synaptic input (constant resting potential $V_\text{rest}$ and common external
stimulus $I_{\text{ext}}(t)$). In the fully-connected network, the synaptic current $I_{\text{syn},i}(t)$ is the same for all neurons $i$ and is given by Eq.~\eqref{eq:I_syn}.

In each time interval $[t,t+dt)$ spikes are drawn with probability $\lambda_i(t)dt$, where the firing rate $\lambda_i(t)$ depends on the input potential as follows:
\begin{equation}
\lambda_i(t)=\Phi\bigl(h_i(t)\bigr). 
\end{equation}
Here, we choose a sigmoidal transfer function of the form $\Phi(h)=r_\text{max}/\left[1+\exp(-\beta(h-h_0))\right]$ \gruen{and $h_0 = 0$ mV}.

\subsubsection{GIF model}
\label{sec:gif-model}
 The GIF model for the postsynaptic neuron dynamics is determined by the membrane potential $V_i(t)$, the dynamic threshold $\vartheta_i(t)$ and a conditional intensity $\lambda_i(t)$ for the stochastic spike generating mechanism. Here, the index $i=1,\dotsc,N$ represents the neuron label. Between spikes, the membrane potential satisfies the dynamics
\begin{equation}
  \label{eq:glif}
  \taum\od{V_i}{t}=-V_i+\mu(t)+R_\text{m}I_{\text{syn},i}(t),
\end{equation}
where the quantities $\taum$, $\mu(t)$, $R_\text{m}$ and $I_{\text{syn},i}(t)$ are the same as in the rate model above.

After a spike, the voltage is immediately reset to the potential $\vreset$, where it is clamped for an absolute refractory period
$\tref=4$~ms. A spike also affects the threshold variable $\vartheta$: if $s_i(t)$ denotes the spike train of the postsynaptic neuron, the dynamics of the threshold can be written as
\begin{equation}
  \label{eq:thresh}
\vartheta_i(t)=\vth+\int_{-\infty}^t\theta(t-t')s_i(t')\,\mathrm{d}t',
\end{equation}
where $\vth$ is a baseline threshold and $\theta(\tau)$ is
spike-triggered threshold kernel \cite{MenNau12,PozMen15}, which is added to the threshold upon each spike. To silence the neuron during the absolute refractory period, we set $\theta(\tau)=\infty$ for $\tau\in[0,\tref)$.

Spikes are generated by a conditional intensity (hazard rate) of an exponential form:
\begin{equation}
  \label{eq:hazard-def}
\lambda_i(t)=c\exp\left(\frac{V_i(t)-\vartheta_i(t)}{\Delta_u}\right).
\end{equation}
This means that the conditional intensity, and hence the probability $\lambda_i(t)dt$ to fire in the interval $[t,t+dt)$, depends on the momentary distance between the membrane potential and threshold. For simplicity, in this paper we use a constant threshold, $\vartheta_i(t) = \vth$, i.e. we do not consider neuronal adaptation. This completes the definition of the microscopic model.

\subsection{Mesoscopic mean-field model}
\label{sec:meso-recurr}

\rot{As explained in \cite{SchDeg17}, the random connectivity can be well approximated by a fully connected network ($C=N$), with rescaled synaptic weights $pw$, corresponding to a mean-field approximation. In the following, we shall therefore choose $p=1$ unless stated otherwise.  In the mean-field approximation, the TPSI $y(t)$ and the synaptic current $I_{\text{syn}}(t)$ do not depend on the identities $j$ of the postsynaptic neurons and are related by Eq.~\eqref{eq:1} with $J=\frac{\taum}{R_\text{m}}Npw$. For the case of exponential synapses, the synaptic current reads
\begin{equation}
	\label{eq:I_syn-mf}
    \tau_s \od{I_{\text{syn}}}{t} = -I_{\text{syn}}+J \hat{u}(t)\hat{x}(t)A(t), 
\end{equation}
where $\hat{u}$ and $\hat{x}$ obey the mean-field equations Eqs.~\eqref{eq:gaussian} -- \eqref{eq:meso-full}.

As shown in \cite{SchDeg17}, the population activity $A(t)$ can be determined by a single mesoscopic variable, the instantaneous rate $r(t)$, whose dynamics depends on the microscopic model and the history of the population activity $\mathcal{H}_t=\{A(t')|t'<t\}$. Specifically, $A(t)$ is given by the normalized spike train }
\begin{equation}
  \label{eq:A-N-meso}
  A(t)=\frac{1}{N}\frac{dn(t)}{dt}=\frac{1}{N}\sum_{k\in \mathbb{Z}}\delta(t-\hat{t}_k),
\end{equation}
where $n(t)$ is a counting process with (conditional) intensity $\hat{\lambda}(t)=Nr(t)$  representing the total number of spikes in the population up to time $t$. The second equality means that $A(t)$ is proportional to a spike train with spike times $\hat{t}_k$ generated with rate $Nr(t)$. This is similar to the superposition of Poisson spike train in the feedforward case, Sec.~\ref{sec:poisson}, where the pooled spike train also exhibits the rate $Nr(t)$.  For large $N$ the finite-size population activity has the simple representation
\begin{equation}
  \label{eq:A-N-meso-poisson}
  A(t)=r(t)+\sqrt{\frac{r(t)}{N}}\xi(t),
\end{equation}
where $\xi(t)$ is a Gaussian white noise with zero mean and correlation function $\langle\xi(t)\xi(t')\rangle=\delta(t-t')$. The Gaussian representation \gruen{is} useful in simulations when \gruen{the number $N$ of neurons in the population} is large.

\subsubsection{Poisson rate model}
In the Poisson rate model, \rot{the rate $r(t)$ is given by
\begin{align}
r(t)&=\Phi\bigl(h(t)\bigr),\\
\taum \od{h}{t} &=-h+\mu(t)+R_{\text{m}} I_{\text{syn}}(t),
\end{align}
Importantly, the coupling to the STP dynamics is contained in the synaptic current $I_\text{syn}$ governed by  Eq.~\eqref{eq:I_syn-mf} and the synaptic mean-field dynamics given by Eqs.~\eqref{eq:gaussian} -- \eqref{eq:meso-full}. 
}

\subsubsection{GIF population model}
\label{sec:gif-population-model}
For the model with spike-history dependence, \rot{the rate $r(t)$ is obtained from an integral over refractory states. A possible representation of the neuronal refractory state is given by the time $\tau$ since the last spike (``age of the neurons''; an alternative representation in terms of the last spike times $\tl=t-\tau$ is given in the appendix, Sec.~\ref{sec:char-equat-param}). Given the distribution of ages in the population, $q(\tau,t)$, the rate at time $t$ results from \cite{SchDeg17}, 
\begin{equation}
  \label{eq:A-bar}
r(t)=\int_0^\infty\lambda(t,\tau)q(\tau,t)\,d\tau+\Lambda(t)\lrrund{1-\int_0^\infty q(\tau,t)\,d\tau},
\end{equation}
where the density $q(\tau,t)$ evolves according to the first-order partial differential equation with time-dependent boundary condition at $\tau=0$:
\begin{equation}
  \label{eq:quasi-lin-q}
  (\partial_t+\partial_\tau)q =-\lambda (t,\tau)q ,\qquad q (0,t)=A (t).
\end{equation}
Here, $A(t)$ is given by Eq.~\eqref{eq:A-N-meso}. In Eqs.~\eqref{eq:A-bar} and \eqref{eq:quasi-lin-q}, the functions $\lambda$ and $\Lambda$ are given by
\begin{equation}
  \label{eq:lam-refr}
\lambda(t,\tau)=c\exp\left(\frac{V(t,\tau)-\vartheta(t,\tau)}{\Delta_u}\right),\qquad
\Lambda(t)=\frac{\int_0^{\infty}\lambda(t,\tau)W(t,\tau)\,\mathrm{d}\tau}{\int_0^{\infty}W(t,\tau)\,\mathrm{d}\tau},
\end{equation}
where $V$ and $W$ and $\vartheta$ are dynamical variables that obey the following dynamics: The age-dependent membrane potential $V(\tau,t)$ and variance function $W(\tau,t)$ follow the first-order partial differential equations
\begin{align}
  \label{eq:uv-quasilin_u-main}
  (\partial_t+\partial_\tau)V &=-\frac{V -\mu }{\taum }+\frac{R_{\text{m}}}{\taum} I_{\text{syn}}(t),\\
  \label{eq:uv-quasilin_v-main}
  (\partial_t+\partial_\tau)W &=-\lambda (t,\tau)[2W -q ]
\end{align}
with  boundary conditions $V(t,0)=\vreset$ and $W(t,0)=0$. The coupling to the STP mean-field dynamics, Eqs.~\eqref{eq:gaussian} -- \eqref{eq:meso-full}, is contained in the synaptic current $I_{\text{syn}}$ governed by  Eq.~\eqref{eq:I_syn-mf}, which influences the voltage $V(t, \tau)$ (Eq.~\eqref{eq:uv-quasilin_u-main}), and hence changes $\lambda(t, \tau)$ and $\Lambda(t)$.

In general, the threshold function $\vartheta$ is given by
\begin{equation}
  \label{eq:thresh-tau}
\vartheta(t,\tau)=\vth+\theta(\tau)+\int_{\tau}^{\infty}\tilde{\theta}(\tau')A(t-\tau')\,\mathrm{d}\tau',
\end{equation}
where $\tilde{\theta}(\tau)=\Delta_u\bigl[1-e^{-\theta(\tau)/\Delta_u}\bigr]$ \cite{NauGer12,SchDeg17}. Note that the last term in Eq.~\eqref{eq:thresh-tau} accounts for spike-history effects before the last spike through the past population activity $A(t')$, $t'<t-\tau$. In this paper, however, Eq.~\eqref{eq:thresh-tau} reduces to a constant value $\vartheta(\tau,t)=\vth$ because we assume a constant threshold $\vartheta(t)=\vth$ in the microscopic model.}

The population equations \eqref{eq:A-bar}--\eqref{eq:uv-quasilin_v} \gruen{has been} be efficiently integrated numerically by the algorithm presented in \cite{SchDeg17}. The numerical integration of the STP mean-field dynamics is given in the Appendix, Sec.~\ref{sec:num_imp}.

\subsection{Recurrent network of Poisson rate neurons -- Microscopic vs. mesoscopic simulations}
\subsubsection{Finite-size noise induced population spikes}
\label{sec:pop_spike}

An interesting example of collective neural dynamics, \gruen{potentially linked to} synaptic depression, \gruen{is the phenomenon of population spikes in cultured neural networks \cite{GigDec15}.} We asked whether population spikes, \gruen{brief period of high average population activity}, can be \gruen{explained} by \gruen{our} finite-size population \gruen{theory with} STP. As in previous work \cite{GigDec15}, we considered a single excitatory population endowed with STP. The mesoscopic mean-field equations allowed us to choose parameters of this model such that the macroscopic mean-field dynamics ($N\rightarrow\infty$, see Eqs.~\eqref{eq:infinite-full}) is \gruen{in an excitable regime for the 2nd-order MF but not for the 1st-order MF.  Here} excitable regime means that the macroscopic dynamics converges to an equilibrium point if \gruen{the total drive} remains below a certain threshold. However, if the threshold is exceeded (e.g. by a brief excitable stimulus or an increase of recurrent synaptic excitation), the activity rises rapidly to large values due to the positive feedback of recurrent excitation. The explosive rise of the activity is terminated by the beginning of synaptic depression, which acts as negative feedback and ultimately wins over recurrent excitation. As a result of the initial excitation, the population activity may show \gruen{population spikes} similar to \gruen{action potential} in other excitable systems such as single neurons. 

\begin{figure*}[t]
  \centering
  \includegraphics[width=\linewidth]{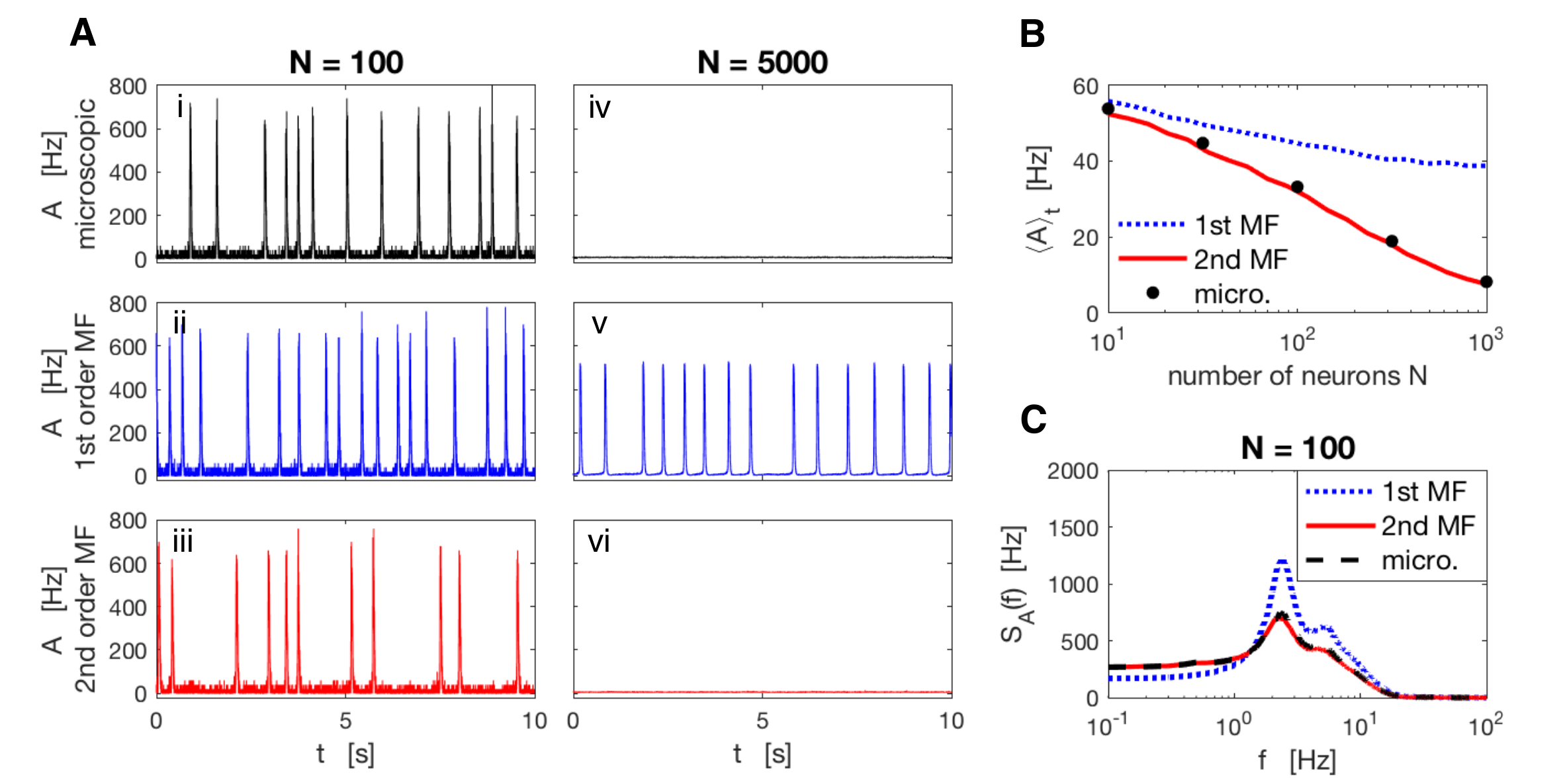}
  \caption{\gruen{\textbf{Recurrent network of Poisson neurons with finite-size noise generating irregular population spikes.} (A) For small population size, $N=100$, microscopic (i) as well as 1st (ii) and 2nd-order (iii) MF dynamics exhibit irregular population spikes. For large population size, $N=5000$, and hence weak finite-size noise,  population spikes cease in microscopic (iv) and 2nd MF dynamics (vi) consistent with an excitable dynamics, whereas the 1st-order MF approximation (v) wrongly predicts regular population spikes corresponding to an underlying oscillatory (limit-cycle) dynamics. (B) Time-averaged population activity decreases with increasing population size indicating a decrease of population spike frequency. The prediction of the 2nd-order MF is accurate across all population sizes, which is not the case with the 1st-order MF, especially for large populations. (C) Power spectra of population activities for $N=100$ neurons. Model parameters are detailed in Appendix~\ref{sec:param}}}
  \label{fig:pop_bif}
\end{figure*}

As expected for an excitable system driven by noise \cite{LinGar04}, the population activity exhibits irregular population spikes if the population size is small (here $N=100$), i.e. if finite-size noise is sufficiently strong  (Fig.~\ref{fig:pop_bif}Ai-iii). In our case, the \gruen{drive} that causes population spikes originates from finite-size fluctuations as expressed by the stochastic terms in Eq.~\eqref{eq:A-N-meso} or \eqref{eq:A-N-meso-poisson}. For $N = 100$, the 2nd-order MF accurately predicts the mean activity (Fig.~\ref{fig:pop_bif}B) and power spectrum (Fig.~\ref{fig:pop_bif}C) of the full microscopic simulation whereas the 1st-order MF deviates quantitatively. Importantly, in the limit of large population size, population spikes vanish in the 2nd-order MF theory consistent with microscopic simulations (Fig.~\ref{fig:pop_bif}Avi,iv, $N=5000$). In marked contrast to microscopic simulations, highly regular population spikes persist even for $N=5000$ in the 1st-order MF approximation corresponding to a determistic limit-cycle dynamics (Fig.~\ref{fig:pop_bif}Av). 

In summary, the 2nd-order MF approximation accurately reproduces the qualitative behavior as well as \gruen{the mean and the power spectrum} of excitatory networks of Poisson neurons with synaptic STP. The statistical properties of the 1st-order MF dynamics exhibit quantitative deviations of statistical properties and \gruen{in some cases} fails to reproduce the qualitative behavior if the system is poised near a bifurcation. The large discrepancies between the 1st-order MF and the microscopic model in the example we show (Fig.~\ref{fig:pop_bif}) are mainly caused by the error in the mean modulation factor $R$. Indeed, for our choice of $\tau_D = \tau_F = 1$~s, correlations between  $u_j$ and $x_j$ are relatively strong but are neglected by the 1st-order MF. We note that this error appears already for the deterministic (i.e. $N\rightarrow\infty$) dynamics. Inaccuracies in the correct description of finite-size noise in the 1st-order MF model may yield additional sources of errors.

\subsubsection{Bistable switching between  Up and Down states induced by finite-size fluctuations}
\begin{figure}[t]
  \centering
  \includegraphics[trim={0.4cm 0 0 0}, scale=.75]{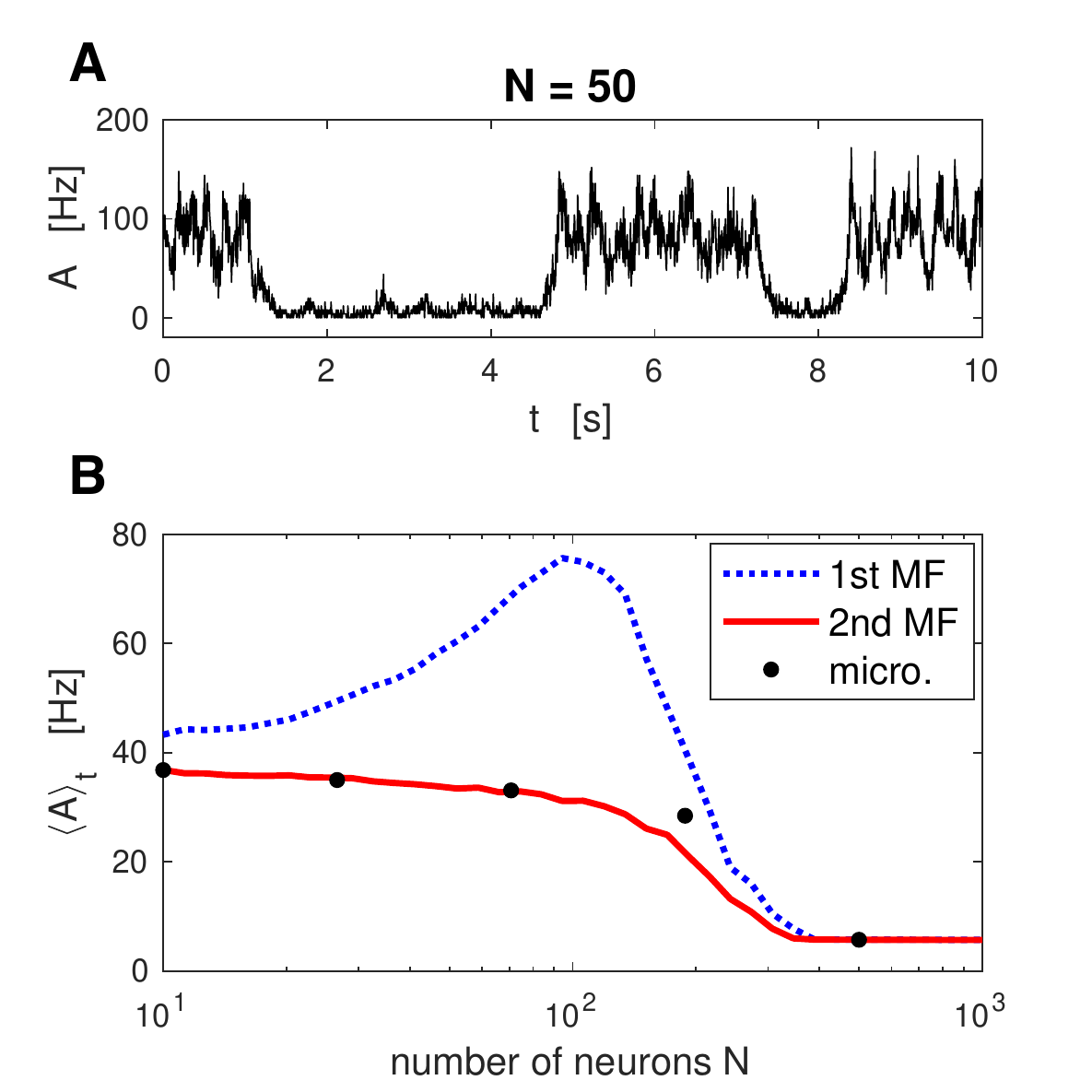}
  \caption{\textbf{Switching between Up and Down states induced by finite-size fluctuations in a recurrent networks of Poisson neurons with STP.} (A) Microscopic simulation of a fully connected network of $N=50$ neurons. The population activity shows switches between Up and Down states. (B) Time-averaged population activity $\langle A \rangle_t$ (which depends on the probabilities of being in the Up and Down states) for different population sizes. Microscopic simulations (black dots) are compared to mesoscopic simulations with the 1st- and 2nd-order MF equations (blue and red lines, respectively). Note that for $N\sim 100$ neurons, the 1st-order MF approximation predicts significantly larger $\langle A \rangle_t$ indicating larger residence times in the Up state due to underestimation of finite-size noise. The same set of parameters are used in (A) and (B) and is detailed in Appendix~\ref{sec:param}.}
  \label{fig:UD}
\end{figure}

Another collective phenomenon in neural networks is multistability. In the presence of finite-size noise, systems with multistable behavior exhibit switches between different attractor states \cite{LitDoi12,MazFon15,SchDeg17}. In particular, bistable neural systems driven by noise support stochastic switches between high and low population activity (``Up and Down states'') \cite{HolTso06,MorRin07,JerRox17,SchDeg17}.  \gruen{As a starting point of our simulations of Up and Down states, following \citet{HolTso06}, we use an excitatory population with synaptic depression in the bistable regime. The qualitative behaviour of the microscopic model exhibiting Up and Down states is captured by both 1st and 2nd-order MF (Fig.~\ref{fig:UD}). A} closer look at the mean firing rate, which is mainly determined by the ratio of the time spent in the Up or Down state, reveals that the 1st-order MF dynamics predicts significantly longer residence times in the Up state (Fig.~\ref{fig:UD}B). In contrast, the 2nd-order MF approximation accuratley matches the simulation of the microscopic model. In this example we have chosen $\tau_D \gg \tau_F$ such that the correlations between $u_j$ and $x_j$ are negligible. As the consequence, the mean modulation factors $R$ predicted by the 1st  and 2nd-order MF theories, and hence the mean TPSI $\langle y\rangle$, are almost equal (cf. Fig.~\ref{fig:params}A). The error made by the 1st-order MF approximation mainly results from an incorrect description of finite-size fluctuations: at high firing rate (Up state), finite-size fluctuations are largely underestimated in the 1st-order MF dynamics as mentionned in Sec.~\ref{sec:tpsi-stat}. The weaker noise implies longer residence times in the Up state. This example highlights the relevance of the fluctuation statistics provided by the 2nd-order MF approximation.


\subsection{Recurrent network of GIF neurons -- Microscopic vs. mesoscopic simulations}
\label{sec:pop-depre}

\begin{figure*}
  \centering
  \includegraphics[width=\linewidth]{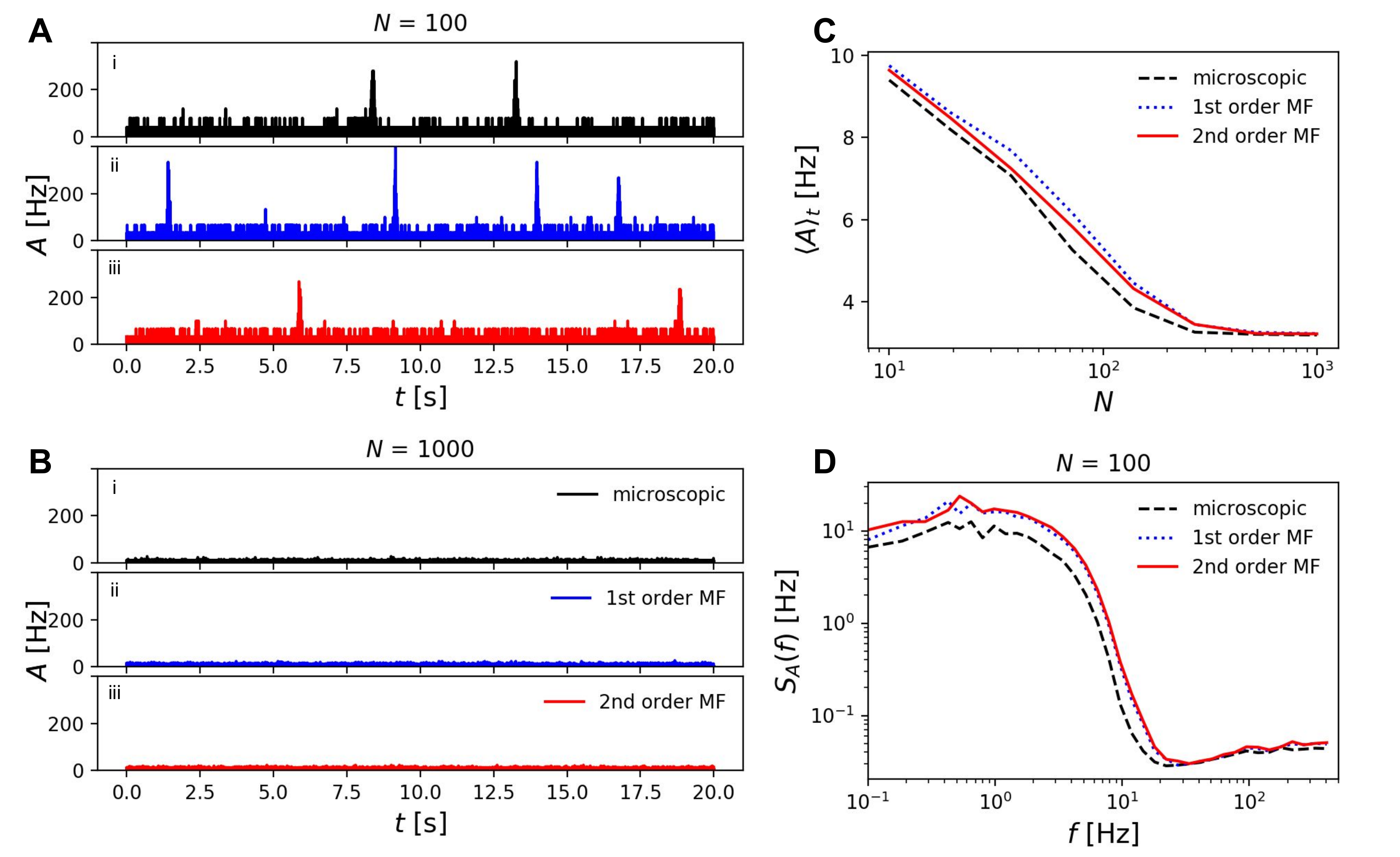}
  \caption{\textbf{ Finite-size noise induced population spikes in recurrent networks of GIF neurons.}  (A) Population activity of a fully connected network of $N=100$ neurons exhibiting irregular population spikes (microscopic simulation (Ai), 1st and 2nd-order MF theory (Aii) and (Aiii), respectively).  (B) Same as (A) but with $N=1000$. Finite-size fluctuations are not strong enough to elicit population spikes. (C) Time-averaged population activity $\langle A \rangle_t$ (which depends on the frequency of population spikes) for different $N$. (D) Power spectral densities of the population activity for a network of 100 neurons. The same set of parameters is used in all panels and is detailed in Appendix~\ref{sec:param}.}
  \label{fig:recurr}
\end{figure*}

As a final demonstration of the mesoscopic MF theory with STP, we consider an excitable regime generating population spikes as in Sec.~\ref{sec:pop_spike} but with more realistic neurons \gruen{described} by a GIF spiking neuron model (Sec.~\ref{sec:gif-model} and \ref{sec:gif-population-model}). Because of spike-history dependencies such as refractoriness, spike arrivals at synapses are no longer Poisson processes. Hence, the Poisson assumption of 1st- and 2nd-order MF theories is formally not fulfilled anymore for recurrent GIF networks. Nevertheless, it is interesting to see whether population spikes can still be captured by the mesoscopic MF equations. To this end, we simulated the recently developped  mesoscopic population equations for populations of GIF neurons \cite{SchDeg17} given by Eqs.~\eqref{eq:I_syn},\eqref{eq:A-bar}--\eqref{eq:uv-quasilin_v} extended by the MF equations for STP, Eqs.\eqref{eq:gaussian} -- \eqref{eq:meso-full}.  The full MF theory qualitatively reproduces population spikes at small population sizes  (Fig.~\ref{fig:recurr}A) and their extinction for large populations (Fig.~\ref{fig:recurr}B). Both the mean (Fig.~\ref{fig:recurr}C) and fluctuation statistics (Fig.~\ref{fig:recurr}D) are roughly captured by the MF equations albeit with small deviations from the microscopic simulation. However, a clear advantage of \gruen{2nd vs. 1st} order approximation is not apparent. This indicates that the 2nd-order approximation does not necessarily yield a better approximation for networks of non-Poisson neurons and that the computationally simpler 1st-order MF model might be preferable in spiking neural networks with strong spike-history effects.

\section{Discussion}
\label{sec:discuss}

We have derived stochastic mean-field (MF) equations that capture the effect of synaptic short-term plasticity (STP) at the level of populations. These equations generalize previous MF theories \gruen{for deterministic population rates} \cite{TsoPaw98,HolTso06,BarTso07} to the case of finite-size populations with stochastic population rates (mesoscopic description). The mesoscopic STP dynamics is compatible with a recent mesoscopic population model \cite{SchDeg17}, which has been originally derived for static synapses. The mesoscopic MF dynamics of STP can thus be easily included into existing mesoscopic models. We find that a first-order mean-field approximation that accounts for stochastic rates but neglects correlations between facilitation and depression variables (as in \citet{TsoMar97})  well accounts for the mean stationary input. This mean input is slightly improved by a second-order approximation, which accounts for correlations but neglects third and higher-order cumulants. The main strength of the 2nd-order MF theory lies in the prediction of fluctuations and transient responses \gruen{of the STP variables}. We have shown that population spikes and UP and Down state switches in a one-population model with synaptic depression can be well described by the extended mesoscopic model. In particular, the 2nd-order MF equations accurately replicate simulations of a network of Poisson neurons coupled via dynamic synapses. For networks of GIF spiking neurons the agreement is less accurate but still captures the qualitative collective dynamics.

In simulations of neuronal populations with STP, our mesoscopic mean-field model yields a considerable reduction of computational complexity. Compared to a network with static synapses, each neuron is endowed with two additional variables $u_j$ and $x_j$ that capture the effect of dynamic synapses onto its postsynaptic target neurons. In a \gruen{single population} of $N$ neurons and connection probability $p$, a microscopic simulation thus requires the numerical integration of $2pN$ additional equations. By contrast, a simulation of the mesoscopic model only needs $4$ additional equations per population. Thus we expect that our extended mesoscopic dynamics offers a significant speed up of large-scale simulations of cortical circuits with dynamic synapses \cite{MarMul15}.

An interesting question that has been studied theoretically \cite{MerLin10,RosRub12,DroSch13}  is how STP affects information transmission through a large ensemble of dynamic synapses. Our reduction of a synaptic ensemble to a four-dimensional nonlinear mean-field dynamics offers a mathematical framework to derive approximate analytical expressions for measures of information transmission. \gruen{Analysing} information processing capabilities of STP \gruen{in the context of our mean field theory} is an interesting topic for future studies.

We have employed the deterministic STP model of Tsodyks and Markram \cite{TsoPaw98}. While the resulting mean-field equations hold for this specific model, the same approach can be applied straightforwardly to other deterministic models of STP (e.g. \citet{DittKre00,DayAbb05}).   
It is less obvious how to treat {\em stochastic} models of STP. Biological synapses are highly stochastic owing to the small number of synaptic vesicles that are randomly released upon spike arrival. This includes a finite probability of transmission failure. Using stochastic models of STP, it has been shown that synaptic stochasticity has a strong impact on information transmission \cite{RosRub12} and postsynaptic neural responses \cite{BirRic14}. On the population level, it seems to be feasible to treat this source of randomness in a similar manner as we did in Sec.~\ref{sec:meso-var}. A generalization to a mesoscopic STP model that is \gruen{applicable for stochastic synapses}, will be an important subject for further studies.

The mean-field equations for the STP dynamics have been derived under the assumption that presynaptic spike trains are (time-inhomogeneous) Poisson processes. We have tested the mean-field equations in a feedforward setup and a recurrent network of Poisson rate units, where the Poisson assumption holds true, and found excellent agreements with microscopic simulations. For the application to recurrent networks of generalized integrate-and-fire (GIF) neurons in Sec.~\ref{sec:pop-depre}, the Poisson assumption is not fulfilled because of refractoriness and other spike-history dependencies of single neurons \cite{DegSch14,SchDro15}. Despite of the non-Poisson (colored-noise) statistics of spike trains in integrate-and-fire networks, a Poisson (white-noise) assumption is commonly used in mean-field theories as a ``first-order approximation'' \cite{Bru00}. In a similar spirit, we here simply assumed that that synaptic input can be treated as a Poisson process so as to apply our MF theory for STP to networks of GIF neurons. For a simple one-population model with excitatory synaptic connections and STP that exhibits nontrivial dynamics in the form of population spikes, we have shown that the MF equations  reproduce the qualitative behavior. This indicates that the MF theory may be valid beyond networks of Poisson neurons. A theoretical analysis of the effect of non-Poisson inputs and the region of validity of the present MF model is beyond the scope of the present paper and remains to be studied. To this end, theoretical approaches to treat dynamic synapses driven by renewal processes \cite{MonHan2012,BirRic2018} might be a promising starting point.

In our previous work \cite{SchDeg17}, we have developed a mean-field theory of neuronal populations that incorporates spike-history dependencies, such as refractoriness and adaptation, and finite-size fluctuations in a consistent manner. By adding another important feature -- synaptic short-term plasticity -- we have here made a further step towards a microscopically-grounded mesoscopic population model of a cortical circuit.

\appendix
\section{Derivation of mesoscopic equations}
\label{sec:meso-eq}

To derive the mesoscopic Eqs.~\eqref{eq:gaussian}--\eqref{eq:meso-full}, it is useful to rewrite the synaptic dynamics in differential form:
\begin{align}
  \label{eq:stp-differential-u}
  du_j&=\frac{U_0-u_j}{\tau_F}dt+U(1-u_j)dn_j(t),\\
  \label{eq:stp-differential-x}
  dx_j&=\frac{1-x_j}{\tau_D}dt-u_jx_jdn_j(t).
\end{align}
Here, $dn_j(t)$ denotes the increment of the Poisson count process
$n_j(t)=\int^ts_j(t')\,dt'$ at the $i$-th synapse in an infinitesimal time interval $[t,t+dt)$. Because of the
Poisson property, the increments $dn_j(t)$ are independent at different
times $t$. Equation~\eqref{eq:stp-differential-u} and
\eqref{eq:stp-differential-x} are interpreted in the Ito sense,
i.e. the variables multiplying $dn_j$ are evaluated before the jumps
of $n_j$. Therefore, $u_j(t)$ as well as $u_j(t)x_j(t)$ are independent of
$dn_j(t)$.

We want to derive differential equations for the mesoscopic variables
$u(t)=N^{-1}\sum_ju_j(t)$ and $x(t)=N^{-1}\sum_jx_j$ only using the mesoscopic spike count $n(t)=\sum_jn_j(t)$. Taking the temporal derivative yields
\begin{align}
  \label{eq:sum-u}
  du&=\frac{U_0-u}{\tau_F}dt+U\left(\frac{dn}{N}-\frac{1}{N}\sum_j u_jdn_j\right),\\
  \label{eq:sum-x}
  dx&=\frac{1-x}{\tau_D}dt-\frac{1}{N}\sum_ju_jx_jdn_j.
\end{align}
The sums over the weighted spike counts $dn_j$ cannot be expressed as a deterministic function of the mesoscopic spike count $dn$. However, we can make use of the fact that in an infinitesimally small time interval $dt$ at most one spike can occur and contribute to the sums, which thereby simplify considerably: $dn_j$ is either zero for all neurons $i$ (i.e. $dn=0$), or there exists one and only one neuron $i$ for which $dn_j=1$ (in this case $dn=1$). This implies that we can write for any function $g(u_j,x_j)$ multiplying $dn_j$
\begin{equation}
  \label{eq:trick}
  \frac{1}{N}\sum_jg(u_j,x_j)dn_j=g(u_j,x_j)\frac{dn}{N},
\end{equation}
where $j$ is the index of the neuron spiking at time $t$ (if no neuron spikes $dn=0$). For independent Poisson processes, the index $j$ is a random number that is uniformly drawn from the set of possible neuron indices $\{1,\dotsc,N\}$. In a stochastic simulation, this picture corresponds to the generation of Poisson realizations by drawing a first random number to decide whether there is some spike in the presynaptic population in the present time interval $[t,t+dt)$, and if this is the case, drawing a second random number which determines the neuron to which the spike should be attached. Because at time $t$ the variables $u_j(t)$, $x_j(t)$ vary independently across different  synapses, the random picking of a synapse $i$ entails that $u_j$ and $x_j$ can be regarded as random numbers $\hat{u}$ and $\hat{x}$ that are drawn at each spike time from a distribution that is consistent with the current distribution of $u_j$ and $x_j$ across the population. Here, we assume a Gaussian distribution $p(u,x,t)$ with mean $(u,x)$ and covariance matrix $\Sigma(t)$ given by \eqref{eq:mus} and Eq.~\eqref{eq:cova}. This allows us to rewrite Eq.~\eqref{eq:trick} as
\begin{equation}
  \label{eq:trick2}
  \frac{1}{N}\sum_jg(u_j,x_j)dn_j=g(\hat{u},\hat{x})\frac{dn}{N}.
\end{equation}
Applying this relation to Eqs.~\eqref{eq:sum-u} and \eqref{eq:sum-x} yields
\begin{subequations}
  \label{eq:u-x-MF}
\begin{align}
  \label{eq:sum-u-mf}
  du&=\frac{U_0-u}{\tau_F}dt+U\left(1-\hat{u}\right)\frac{dn}{N},\\
  \label{eq:sum-x-mf}
  dx&=\frac{1-x}{\tau_D}dt-\hat{u}\hat{x}\frac{dn}{N}.
\end{align}
\end{subequations}
Using the definition of the population activity $A(t)=dn(t)/(Ndt)$ yields Eqs.~\eqref{eq:meso-u} and \eqref{eq:meso-x} in the main text. 

In the covariance matrix, Eq.~\eqref{eq:cova}, the mesoscopic quantities $P=N^{-1}\sum_ju_j^2$, $Q=N^{-1}\sum_jx_j^2$ and
$R=N^{-1}\sum_ju_jx_j$ appear. We now derive stochastic
differential equation for these variable. For instance, for the total differential $dP$ we need to evaluate $d(u_j^2)$:%
\begin{align}
  d\bigl(u_j^2\bigr)&=u_j^2(t+dt)-u_j^2(t)=(u_j+du_j)^2-u_j^2\nonumber\\
    &=2u_jdu_j+du_j^2.  \label{eq:P-deriv}
\end{align}

Similarly, we write $d(x_j^2)=2x_jdx_j+dx_j^2$ and $ d(u_jx_j)=(u_j+du_j)(x_j+dx_j)-u_jx_j=u_jdx_j+x_jdu_j+du_jdx_j$.

Next, we apply the differential equations \eqref{eq:stp-differential-u} and
\eqref{eq:stp-differential-x} for $u_j$ and $x_j$. By doing so, we use Ito's
calculus and the property that $dn_j$ is of order $\sqrt{dt}$ and that
$dn_j^k=dn_j$ for all integer powers $k\ge 1$. These properties result from the fact that in a small time step $dt$, $dn_j$ can be seen as Bernoulli variable with values $0$ or $1$ and variance $r(t)dt$. As it is commonly done in derivations of stochastic differential equations, we will keep terms of order $\sqrt{dt}$ and $dt$ but discard higher-order terms proportional to $dtdn_j$ and $dt^2$. For the nonlinear terms in Eq.~\eqref{eq:P-deriv} we thus obtain
\begin{align}
  \label{eq:duisquare}
  (du_j)^2&=U^2(1-u_j)^2dn_j,\\
  u_jdu_j&=u_j\frac{U_0-u_j}{\tau_F}dt+Uu_j(1-u_j)dn_j.
\end{align}
Similarly, we find
\begin{align}
  \label{eq:dxisquare}
  (dx_j)^2&=u_j^2x_j^2dn_j,\\
  x_jdx_j&=x_j\frac{1-x_j}{\tau_D}dt-u_jx_j^2dn_j,\\
  u_jdx_j&=u_j\frac{1-x_j}{\tau_D}dt-u_j^2x_jdn_j,\\
x_jdu_j&=x_j\frac{U_0-u_j}{\tau_F}dt+Ux_j(1-u_j)dn_j,\\
  du_jdx_j&=-Uu_jx_j(1-u_j)dn_j.
\end{align}
Summing the differentials $d(u_j^2)$, $d(x_j^2)$ and $d(u_jx_j)$ over all $N$ synapses and using Eq.~\eqref{eq:trick2}, we obtain
\begin{subequations}
\label{eq:PQR-appendix}
\begin{align}
  \label{eq:dK}
  dP&=2\frac{U_0u-\muuu}{\tau_F}dt+U(1-\hat{u})[(2-U)\hat{u}+U]\frac{dn}{N},\\
  dQ&=2\frac{x-\muxx}{\tau_D}dt-\hat{u}\hat{x}^2(2-\hat{u})\frac{dn}{N}\\
dR&=\lrrund{\frac{U_0x-\muux}{\tau_F}+\frac{u-\muux}{\tau_D}}dt+\hat{x}\left[U(1-\hat{u})^2-\hat{u}^2\right]\frac{dn}{N}
\end{align}
\end{subequations}
These equations correspond to Eqs.~\eqref{eq:meso-P}, \eqref{eq:meso-Q} and \eqref{eq:meso-R} in the main text.

\section{Efficient numerical implementation of mesoscopic equations}
\label{sec:num_imp}

The mesoscopic STP dynamics given by Eqs.~\eqref{eq:meso-full} or Eqs \eqref{eq:u-x-MF} and \eqref{eq:PQR-appendix} are driven by a point process $A(t)$ or increments $dn(t)$ that are multiplied by a stochastic factor of the form $f(\hat{u},\hat{x})$. In simulations, these stochastic amplitudes require some care. A straightforward discretization of Eqs \eqref{eq:u-x-MF} and \eqref{eq:PQR-appendix} would be to draw $\hat{u}$ and $\hat{x}$ from their joint distribution in each time step independently, compute  $f(\hat{u},\hat{x})$ and multiply by the number of spikes
\begin{equation}
  \label{eq:Deltan}
  \Delta n(t)=\int_t^{t+\Delta t}dn(t')=N\int_t^{t+\Delta t}A(t')\,dt'
\end{equation} 
that occur in the discretization interval $[t,t+\Delta t)$. However, this approach is only correct if the discretization time step $\Delta t$ is small enough such that $\Delta n$ contains at most one spike. Because spikes result from a population of many neurons, this condition would require an extremely small time step (such that $Nr(t)\Delta t\ll 1$), and would thus yield a highly inefficient simulation algorithm. Luckily, the independence of the factors  $f(\hat{u},\hat{x})$ across spikes in the interval $[t,t+\Delta t)$ allows us to use larger time steps that may contain multiple spikes\footnote{Here we assume that different spikes in $[t,t+\Delta t)$ belong to different neurons, which is justified if $r(t)\Delta t\ll 1$.}: due to the independence the integration of the stochastic term in Eqs \eqref{eq:u-x-MF} and \eqref{eq:PQR-appendix} simplifies to a sum of $\Delta n$ {\it i.i.d.} random variables:
\begin{equation}
  \label{eq:dn-form}
  \int_t^{t+\Delta t}f(\hat{u},\hat{x})dn(t')=\sum_{j=1}^{\Delta n(t)}f(\hat{u}_j,\hat{x}_j).
\end{equation}
If the random variables $f(\hat{u}_j,\hat{x}_j)$ have mean $\mu_f$ and variance $v_f$, the sum in Eq.~\eqref{eq:dn-form} has mean $\mu_f\Delta n$ and variance $v_f\Delta n$. Using a Gaussian approximation, we can thus approximate the integral, Eq.~\eqref{eq:dn-form}, by
\begin{equation}
  \label{eq:stoch_int}
  \int_t^{t+\Delta t}f(\hat{u},\hat{x})dn(t')\approx \mu_f\Delta n(t)+\sqrt{\Delta n(t)}\varepsilon(t),
\end{equation}
where $\varepsilon(t)$ is a centered Gaussian random variable with variance $v_f$ that is independent in each time step.

 For example, the integration of the mesoscopic equation for $u$, Eq.~\eqref{eq:sum-u-mf}, involves the integration of the stochastic term
 $N^{-1}U(1-\hat{u})dn$. Using $\langle\hat{u}\rangle=\mu_u=u$ and $\var{\hat{u}}=v_u=P-u^2$,  Eq.~\eqref{eq:stoch_int} yields
\begin{equation*}
\frac{U}{N}\int_t^{t+\Delta t}[1-\hat{u}(t')]dn(t')\approx \frac{U}{N}\lreckig{(1-u)\Delta n-\varepsilon_u(t)\sqrt{\Delta n} },
\end{equation*}
where $\varepsilon_u(t)$ is a centered Gaussian random variable with variance $v_u$. Similarly, the integration of the mesoscopic equation for $x$, Eq.~\eqref{eq:sum-x-mf}, yields
\begin{multline}
\int_t^{t+\Delta t}\hat{u}(t')\hat{x}(t')dn(t')\\
=\int_t^{t+\Delta t}(u(t')+\varepsilon_u(t'))(x(t')+\varepsilon_x(t'))dn(t')\\
=\int_t^{t+\Delta t}[u(t')x(t')+\varepsilon_u(t')\varepsilon_x(t')]dn(t')\\
+\int_t^{t+\Delta t}[\varepsilon_u(t')x(t')+\varepsilon_x(t')u(t')]\,dn(t')\\
\approx R(t)\Delta n(t)+[x(t)\varepsilon_u(t)+u(t)\varepsilon_x(t)]\sqrt{\Delta n(t)}.
\end{multline}
Here, we approximated $\varepsilon_u\varepsilon_x$ by its mean $\langle\varepsilon_u\varepsilon_x\rangle=R-ux$, and neglected its fluctuations. 

When we integrate Eqs.~\eqref{eq:PQR-appendix}, we encounter the terms $\varepsilon_u^2 \varepsilon_x$, $\varepsilon_u \varepsilon_x^2$ and $\varepsilon_u^2 \varepsilon_x^2$. As we cannot calculate their mean, we perform a moment closure approximation, neglecting all cumulants of order higher than two:
\begin{equation*}
  \langle \varepsilon_u^2 \varepsilon_x \rangle \approx 2\langle \varepsilon_u \rangle \langle \varepsilon_u \varepsilon_x\rangle + \langle \varepsilon_u^2 \rangle \langle \varepsilon_x \rangle -2\langle \varepsilon_u\rangle^2 \langle\varepsilon_x\rangle = 0;
\end{equation*}
symmetrically, $\langle \varepsilon_u \varepsilon_x^2 \rangle \approx 0$; and finally,
\begin{align*}
  \langle \varepsilon_u^2 \varepsilon_x^2 \rangle &\approx \langle \varepsilon_u^2 \rangle \langle \varepsilon_x^2 \rangle + 2\langle \varepsilon_u \varepsilon_x \rangle^2 \\
  &= (P-u^2)(Q-x^2) + 2(R-ux)^2.
\end{align*}
These approximations allow for the completion of the derivation. 

In summary, the Euler scheme corresponding to Eqs.~\eqref{eq:meso-full} or, respectively, Eqs \eqref{eq:u-x-MF} and \eqref{eq:PQR-appendix},  is:
  \begin{gather*}
    \label{eq:update}
    u(t+\Delta t)=u(t)+\Delta u,\quad x(t+\Delta t)=x(t)+\Delta x,\\
    P(t+\Delta t)=P(t)+\Delta P,\quad Q(t+\Delta t)=Q(t)+\Delta Q,\\
    R(t+\Delta t)=R(t)+\Delta R
  \end{gather*}
with increments given by
\begin{subequations}
\begin{align}
  \Delta u &=\frac{U_0-u}{\tau_F}\Delta t+\frac{U}{N}\left[(1-u)\Delta n - \varepsilon_u\sqrt{\Delta n}\right], \\
\Delta x &=\frac{1-x}{\tau_D}\Delta t-\frac{1}{N}\left[R\Delta n +(u\varepsilon_x + x\varepsilon_u)\sqrt{\Delta n}\right], \\
  \Delta P &=2 \frac{U_0u-\muuu}{\tau_F}\Delta t+\frac{1}{N}\lreckig{\mu_P\Delta n +\varepsilon_P\sqrt{\Delta n}},\\
 \Delta Q &=2 \frac{x-\muxx}{\tau_D}\Delta t+\frac{1}{N}\lreckig{\mu_Q\Delta n +\varepsilon_Q\sqrt{\Delta n}},\\
\Delta R &= \frac{U_0x-\muux}{\tau_F}\Delta t+\frac{u-\muux}{\tau_D}\Delta t+\frac{1}{N}\lreckig{\mu_R\Delta n +\varepsilon_R\sqrt{\Delta n}}.
\end{align}
\end{subequations}
Here, we abbreviated
\begin{align*}
  \mu_P &= U \big(P (U - 2) - 2 u (U - 1) + U\big),\\
  \varepsilon_P &= 2U \big(1 + u (U - 2) - U\big)\varepsilon_u, \\
  \mu_Q &= P Q - 2 Q u + 2 \big(R + (u - 2) x\big) (R - u x),\\
  \varepsilon_Q &= 2(u - 1)x^2\varepsilon_u + 2u(u-2)x\varepsilon_x,\\
  \mu_R &= (U (1 - u)^2 - u^2) x + (U - 1) x (P - u^2) \\
           & \quad + 2 (U (u - 1) - u) (R - u x),\\
  \varepsilon_R &= 2\big(U(u - 1) - u\big)x\varepsilon_u + \big(U(1 - u)^2 - u^2\big)\varepsilon_x.
\end{align*}

The generatation of correlated Gaussian random variables with covariance matrix given by Eq.~\eqref{eq:cova} can be implemented by standard methods. For instance, one may compute in each time step the correlation coefficient
\begin{equation}
  \label{eq:corrcoeff}
  \rho=\frac{R-ux}{\sqrt{(P-u^2)(Q-x^2)}}.
\end{equation}
Initially, it may happen that the numerical values of the variances $P-u^2$ and $Q-x^2$ are non-positive or the absolute value of the correlation coefficient $|\rho|$ exceeds unity. In these cases, we set $\varepsilon_u$ and $\varepsilon_x$ to zero. Otherwise, we generate random variates by the formula
\begin{align}
  \label{eq:randomvars}
  \varepsilon_u&=\sqrt{P-u^2}z_1,\\
  \varepsilon_x&=\sqrt{Q-x^2}\lrrund{\rho z_1+\sqrt{1-\rho^2}z_2},
\end{align}
where $z_1$ and $z_2$ are independent standard normal random numbers.

Finally, we note that in numerical simulations it is convinient to operate on the the spike counts $\Delta n(t)$ rather than on the population activity $A(t)$. The discretized population activity can be easily obtained from the spike counts via the formula
\begin{equation}
  \label{eq:AN-discr}
  A(t)=\frac{\Delta n(t)}{N\Delta t}.
\end{equation}

\section{Equations for infinite-size populations and steady-state formulas}
A macroscopic theory of STP for infinite-size populations for what we call the 1st-order MF has been presented in \cite{TsoPaw98}. We detail here the adaptation of our 2nd-order MF to the case if infinite-size populations. 

In the infinite-size case, the stochastic populations activity $A(t)$ becomes a deterministic rate $r(t)$, which simplifies our mesoscopic equations~\eqref{eq:meso-full}: in the stochastic ODEs, $r(t)$ -- being deterministic -- is not multiplied by a random term but by the mean of this term, transforming the stochahstic ODEs into deterministic ODEs. These means have already been computed in Sec.~\ref{sec:num_imp}. Hence, in the infinite-size case, our 2nd-order MF equations are:
\begin{subequations}
\label{eq:infinite-full}
\begin{align}
  \od{u}{t}&=\frac{U_0-u}{\tau_F}+U(1 - u)r(t),\\
  \od{x}{t}&=\frac{1-x}{\tau_D}-R r(t), \\
  \od{\muuu}{t}&=2\frac{U_0u-\muuu}{\tau_F}+ \mu_P r(t),\\
  \od{\muxx}{t}&=2\frac{x-\muxx}{\tau_D}+ \mu_Q r(t),\\
  \od{\muux}{t}&=\frac{U_0x-\muux}{\tau_F}+\frac{u-\muux}{\tau_D}+ \mu_R r(t).
\end{align}
\end{subequations}
Again, we abbreviated
\begin{align*}
  \mu_P &= U \big(P (U - 2) - 2 u (U - 1) + U\big),\\
  \mu_Q &= P Q - 2 Q u + 2 \big(R + (u - 2) x\big) (R - u x),\\
  \mu_R &= (U (1 - u)^2 - u^2) x + (U - 1) x (P - u^2) \\
           & \quad + 2 (U (u - 1) - u) (R - u x).
\end{align*}Given a fixed rate $r$, we can compute the steady-state values of $u$, $x$, $P$, $Q$ and $R$:
\begin{subequations}
\label{eq:steady-state}
\begin{align}
  u &= \frac{\tau_F r U+U_0}{\tau_F r U+1} \\
  P &= \frac{\tau_F r U (2 u (U-1)-U)-2 u U_0}{\tau_F r (U-2) U - 2} \\
  x &= \frac{\tau_D \tau_F r (-2 u U+u+2 U)+\tau_D+\tau_F}{Z} \\
  R &= \frac{\tau_D \left(\tau_F r \left(P (U-1)-2 u^2 (U-1)+U\right)+U_0\right)+\tau_F u}{Z} \\
  Q &= \frac{-2 \tau_D r (R+(u-2) x) (R-u x)-2 x}{\tau_D r (P-2 u)-2}
\end{align}
\end{subequations}
With the abbreviation
\begin{align*}
  Z &= \tau_D^2 r \left(\tau_F r \left(P (U-1)-2 u^2 (U-1)+U\right)+U_0\right) \\
  &\quad +2 \tau_D \tau_F r (-u U+u+U)+\tau_D+\tau_F.
\end{align*}
Note that these steady-states formulas are very useful for carrying out precise phase-plane analyses.

\rot{
\section{Mesoscopic population equations for network of GIF neurons -- ODE representation}
\label{sec:char-equat-param}

The equation for the population rate $r(t)$ for GIF neurons involves the integral, Eq.~\eqref{eq:A-bar}, over all possible refractory states and a set of partial differential equations (so-called quasi-linear equations) for the quantities $q(\tau,t)$, $V(\tau,t)$ and $W(\tau,t)$. Instead of the age of the neuron (i.e. the time {\em since} its last spike), the refractory state can be equivalently specified by the last spike time $\tl=t-\tau$. This variable transformation turns the partial differential equations into ordinary differential equations (ODEs), which yields an alternative formulation of the mesoscopic population equations.

If the refractory state is specified by the last spike time $\tl=t-\tau$, we need to consider the density of last spike times $Q(\tl,t)\equiv q(t-\tl,t)$. Instead of $Q$, it is slightly more convinient to write $Q(\tl,t)=S(t|\tl)A(\tl)$, where we introduced the survivor function $S(t|\tl)$. With this notation 
the population rate, Eq.~\eqref{eq:A-bar}, becomes \cite{SchDeg17}
\begin{equation}
  \label{eq:r-tlast}
  r(t)=\int_{-\infty}^t\lambda(t|\tl)S(t|\tl)A(\tl)\,d\tl+\Lambda(t)\lrrund{1-\int_{-\infty}^tS(t|\tl)A(\tl)\,d\tl}.
\end{equation}
Using the method of characteristics, the quasilinear equation \eqref{eq:quasi-lin-q} for $q(\tau,t)$ has an equivalent ODE representation for the characteristic curves $\{Q(\tl,t)\}_{t>\tl}$: $dQ/dt=-\lambda Q$ with initial condition $Q(\tl,\tl)=A(\tl)$. This corresponds to an ODE for the survivor function:
\begin{equation}
  \label{eq:survivor}
  \od{S(t|\tl)}{t}=-\lambda(t|\tl)S(t|\tl),\qquad S(\tl|\tl)=1.
\end{equation}
The functions $\lambda(t|\tl)\equiv \lambda(t-\tl,t)$ and $\Lambda(t)$ follow from Eq.~\eqref{eq:lam-refr} as
\begin{equation}
  \label{eq:lam-refr-append}
\lambda(t|\tl)=c\exp\left(\frac{V(t|\tl)-\vartheta(t|\tl)}{\Delta_u}\right),\qquad
\Lambda(t)=\frac{\int_{-\infty}^{t}\lambda(t|\tl)W(t|\tl)\,\mathrm{d}\tl}{\int_{-\infty}^{t}W(t|\tl)\,\mathrm{d}\tl}.
\end{equation}
The dynamics for the variables  $V(t|\tl)$ and $W(t|\tl)$ follow from those of the quasilinear equations \eqref{eq:uv-quasilin_u-main} and \eqref{eq:uv-quasilin_v-main}:
\begin{align}
  \label{eq:uv-quasilin_u}
  \od{V(t|\tl)}{t} &=-\frac{V(t|\tl) -\mu }{\taum }+\frac{R_{\text{m}}}{\taum} I_{\text{syn}}(t)\\
  \label{eq:uv-quasilin_v}
  \od{W(t|\tl)}{t} &=-\lambda (t|\tl)[2W(t|\tl) -S(t|\tl)A(\tl) ]
\end{align}
with  initial conditions $V(\tl|\tl)=\vreset$ and $W(\tl|\tl)=0$. Finally, the threshold function $\vartheta(t|\tl)$ reads
\begin{equation}
  \label{eq:thresh-tau}
  \vartheta(t|\tl)=\vth+\theta(t-\tl)+\int_{-\infty}^{\tl}\tilde{\theta}(t-t')A(t')\,\mathrm{d}t'.
\end{equation}

}

\section{Recurrent Networks parameters}
\label{sec:param}

In Figs.~\ref{fig:pop_bif} and \ref{fig:UD} the following  parameters have been used.

\begin{center}
\begin{tabular}{ |c|c|c|c|c| } 
 \hline
  parameter & unit  & Fig.~\ref{fig:pop_bif} & Fig.~\ref{fig:UD} \\ 
  \hline
 $\tau_D$ & s  & 0.2 & 0.05\\ 
 $\tau_F$ & s  & 0.2 & 0.002\\
 $U$ &  & 0.2 & 0.6\\
 $U_0$ &  & 0.2 & 0.6\\
 $\taum$ & s  & 0.02 & 0.02\\
 $\tau_s$ & s  & 0.002 & 0.01\\
 $\mu$ & mV  & -13.3 & -8.7\\
 $R_{\text{m}}$ & $\mathrm{\Omega}$   & 1 & 1\\
 $\beta$ &  & 0.4 & 0.5\\
 $r_{\text{max}}$ & Hz  & 500 & 200\\
 $h_0$ & mV  & 0 & 0\\
 $J \cdot N$ &  & 78.5 & 28.0\\
 \hline
\end{tabular}
\end{center}
In Fig.~\ref{fig:recurr}, we used the following parameters: $\tau_D$ = 0.5 s, $\tau_F$ =  0.01 s, $U$ = 0.1, $U_0$ = 0.1, $\taum$ = 0.015 s,  $\tau_s$ = 0.01 s, $c$ = 15 Hz, $V_{\text{reset}}$ = 0 mV, $V_{\text{th}}$ = 15 mV, $\mu$ = 4 mV, $R_{\text{m}}$ = 1 $\mathrm{\Omega}$, $\Delta_u$ = 5 mV, $t_{\text{ref}}$ = 0.004 s, $J\cdot N$ = 850.


\begin{thebibliography}{47}
\providecommand{\natexlab}[1]{#1}
\providecommand{\url}[1]{{#1}}
\providecommand{\urlprefix}{URL }
\expandafter\ifx\csname urlstyle\endcsname\relax
  \providecommand{\doi}[1]{DOI~\discretionary{}{}{}#1}\else
  \providecommand{\doi}{DOI~\discretionary{}{}{}\begingroup
  \urlstyle{rm}\Url}\fi
\providecommand{\eprint}[2][]{\url{#2}}

\bibitem[{Abbott et~al(1997)Abbott, Varela, Sen, and Nelson}]{AbbVar97}
Abbott LF, Varela JA, Sen K, Nelson SB (1997) Synaptic depression and cortical
  gain control. Science 275:220

\bibitem[{Barak and Tsodyks(2007)}]{BarTso07}
Barak O, Tsodyks M (2007) Persistent activity in neural networks with dynamic
  synapses. PLoS Comput Biol 3(2):e35, \doi{10.1371/journal.pcbi.0030035}

\bibitem[{Ben-Yishai et~al(1995)Ben-Yishai, Bar-Or, and Sompolinsky}]{BenBar95}
Ben-Yishai R, Bar-Or RL, Sompolinsky H (1995) Theory of orientation tuning in
  visual cortex. Proc Natl Acad Sci USA 92(9):3844--3848

\bibitem[{Bird and Richardson(2014)}]{BirRic14}
Bird A, Richardson M (2014) Long-term plasticity determines the postsynaptic
  response to correlated afferents with multivesicular short-term synaptic
  depression. Frontiers in Computational Neuroscience 8:2,
  \doi{10.3389/fncom.2014.00002}

\bibitem[{Bird and Richardson(2018)}]{BirRic2018}
Bird AD, Richardson MJE (2018) Transmission of temporally correlated spike
  trains through synapses with short-term depression. PLOS Computational
  Biology 14(6):1--25, \doi{10.1371/journal.pcbi.1006232},
  \urlprefix\url{https://doi.org/10.1371/journal.pcbi.1006232}

\bibitem[{Brunel(2000)}]{Bru00}
Brunel N (2000) Sparsely connected networks of spiking neurons. J Comput
  Neurosci 8:183

\bibitem[{Cook et~al(2003)Cook, Schwindt, Grande, and Spain}]{CooSch03}
Cook DL, Schwindt PC, Grande LA, Spain WJ (2003) Synaptic depression in the
  localization of sound. Nature 421(6918):66

\bibitem[{Dayan and Abbott(2005)}]{DayAbb05}
Dayan P, Abbott LF (2005) Theoretical Neuroscience: Computational and
  Mathematical Modeling of Neural Systems, 1st edn. The {MIT} Press

\bibitem[{Deger et~al(2014)Deger, Schwalger, Naud, and Gerstner}]{DegSch14}
Deger M, Schwalger T, Naud R, Gerstner W (2014) Fluctuations and information
  filtering in coupled populations of spiking neurons with adaptation. Phys Rev
  E 90(6-1):062,704, \doi{10.1103/PhysRevE.90.062704}

\bibitem[{Dittman et~al(2000)Dittman, Kreitzer, and Regehr}]{DittKre00}
Dittman JS, Kreitzer AC, Regehr WG (2000) Interplay between facilitation,
  depression, and residual calcium at three presynaptic terminals. J Neurosci
  20:1374

\bibitem[{Droste et~al(2013)Droste, Schwalger, and Lindner}]{DroSch13}
Droste F, Schwalger T, Lindner B (2013) Interplay of two signals in a neuron
  with heterogeneous synaptic short-term plasticity. Front Comp Neurosci 7:86,
  \doi{10.3389/fncom.2013.00086}

\bibitem[{Fiebig and Lansner(2016)}]{FieLan16}
Fiebig F, Lansner A (2016) A spiking working memory model based on hebbian
  short-term potentiation. J Neurosci pp 1989--16

\bibitem[{Fortune and Rose(2001)}]{ForRos01}
Fortune ES, Rose GJ (2001) Short-term synaptic plasticity as a temporal filter.
  Trends Neurosci 24:381

\bibitem[{Gigante et~al(2015)Gigante, Deco, Marom, and Del~Giudice}]{GigDec15}
Gigante G, Deco G, Marom S, Del~Giudice P (2015) Network events on multiple
  space and time scales in cultured neural networks and in a stochastic rate
  model. PLoS Comput Biol 11(11):e1004,547

\bibitem[{Higley and Contreras(2006)}]{HigCon06}
Higley MJ, Contreras D (2006) Balanced excitation and inhibition determine
  spike timing during frequency adaptation. J Neurosci 26(2):448--457

\bibitem[{Holcman and Tsodyks(2006)}]{HolTso06}
Holcman D, Tsodyks M (2006) The emergence of up and down states in cortical
  networks. PLoS Comput Biol 2(3):e23

\bibitem[{Izhikevich and Edelman(2008)}]{IzhEde08}
Izhikevich EM, Edelman GM (2008) Large-scale model of mammalian thalamocortical
  systems. Proc Natl Acad Sci U S A 105(9):3593--3598

\bibitem[{Jercog et~al(2017)Jercog, Roxin, Barthó, Luczak, Compte, and de~la
  Rocha}]{JerRox17}
Jercog D, Roxin A, Barthó P, Luczak A, Compte A, de~la Rocha J (2017) Up-down
  cortical dynamics reflect state transitions in a bistable network. eLife
  6:e22,425, \doi{10.7554/eLife.22425}

\bibitem[{Lefort et~al(2009)Lefort, Tomm, Sarria, and Petersen}]{LefTom09}
Lefort S, Tomm C, Sarria JCF, Petersen CCH (2009) The excitatory neuronal
  network of the c2 barrel column in mouse primary somatosensory cortex. Neuron
  61(2):301--316

\bibitem[{Levina et~al(2007)Levina, Herrmann, and Geisel}]{LevHer07}
Levina A, Herrmann JM, Geisel T (2007) Dynamical synapses causing
  self-organized criticality in neural networks. Nat Phys 3(12):857

\bibitem[{Lindner et~al(2004)Lindner, Garc\'{\i}a-Ojalvo, Neiman, and
  Schimansky-Geier}]{LinGar04}
Lindner B, Garc\'{\i}a-Ojalvo J, Neiman A, Schimansky-Geier L (2004) Effects of
  noise in excitable systems. Phys Rep 392:321

\bibitem[{Lindner et~al(2009)Lindner, Gangloff, Longtin, and Lewis}]{LinGan09}
Lindner B, Gangloff D, Longtin A, Lewis JE (2009) Broadband coding with dynamic
  synapses. J Neurosci 29(7):2076--2088

\bibitem[{Litwin-Kumar and Doiron(2012)}]{LitDoi12}
Litwin-Kumar A, Doiron B (2012) Slow dynamics and high variability in balanced
  cortical networks with clustered connections. Nat Neurosci 15(11):1498--1505

\bibitem[{Markram et~al(1998)Markram, Wang, and Tsodyks}]{MarWan98}
Markram H, Wang Y, Tsodyks M (1998) Differential signaling via the same axon of
  neocortical pyramidal neurons. Proc Natl Acad Sci USA 95(9):5323--5328

\bibitem[{Markram et~al(2015)Markram, Muller, Ramaswamy, Reimann, Abdellah,
  Sanchez, Ailamaki, Alonso-Nanclares, Antille, Arsever et~al}]{MarMul15}
Markram H, Muller E, Ramaswamy S, Reimann MW, Abdellah M, Sanchez CA, Ailamaki
  A, Alonso-Nanclares L, Antille N, Arsever S, et~al (2015) Reconstruction and
  simulation of neocortical microcircuitry. Cell 163(2):456--492

\bibitem[{Mazzucato et~al(2015)Mazzucato, Fontanini, and La~Camera}]{MazFon15}
Mazzucato L, Fontanini A, La~Camera G (2015) Dynamics of multistable states
  during ongoing and evoked cortical activity. J Neurosci 35(21):8214--8231

\bibitem[{Mensi et~al(2012)Mensi, Naud, Pozzorini, Avermann, Petersen, and
  Gerstner}]{MenNau12}
Mensi S, Naud R, Pozzorini C, Avermann M, Petersen CCH, Gerstner W (2012)
  Parameter extraction and classification of three cortical neuron types
  reveals two distinct adaptation mechanisms. J Neurophysiol

\bibitem[{Merkel and Lindner(2010)}]{MerLin10}
Merkel M, Lindner B (2010) Synaptic filtering of rate-coded information. Phys
  Rev E 81(4 Pt 1):041,921--041,921

\bibitem[{Mongillo et~al(2008)Mongillo, Barak, and Tsodyks}]{MonBar08}
Mongillo G, Barak O, Tsodyks M (2008) Synaptic theory of working memory.
  Science 319(5869):1543--1546, \doi{10.1126/science.1150769}

\bibitem[{Mongillo et~al(2012)Mongillo, Hansel, and van Vreeswijk}]{MonHan2012}
Mongillo G, Hansel D, van Vreeswijk C (2012) Bistability and spatiotemporal
  irregularity in neuronal networks with nonlinear synaptic transmission. Phys
  Rev Lett 108:158,101, \doi{10.1103/PhysRevLett.108.158101},
  \urlprefix\url{https://link.aps.org/doi/10.1103/PhysRevLett.108.158101}

\bibitem[{Moreno-Bote et~al(2007)Moreno-Bote, Rinzel, and Rubin}]{MorRin07}
Moreno-Bote R, Rinzel J, Rubin N (2007) Noise-induced alternations in an
  attractor network model of perceptual bistability. J Neurophysiol
  98(3):1125--1139

\bibitem[{Naud and Gerstner(2012)}]{NauGer12}
Naud R, Gerstner W (2012) Coding and decoding with adapting neurons: a
  population approach to the peri-stimulus time histogram. PLoS Comput Biol
  8(10)

\bibitem[{Oswald and Urban(2012)}]{OswUrb12}
Oswald AM, Urban NN (2012) Interactions between behaviorally relevant rhythms
  and synaptic plasticity alter coding in the piriform cortex. J Neurosci
  32(18):6092--6104

\bibitem[{Pittorino et~al(2017)Pittorino, Ib{\'a}{\~n}ez-Berganza, di~Volo,
  Vezzani, and Burioni}]{PitIba17}
Pittorino F, Ib{\'a}{\~n}ez-Berganza M, di~Volo M, Vezzani A, Burioni R (2017)
  Chaos and correlated avalanches in excitatory neural networks with synaptic
  plasticity. Phys Rev Lett 118(9):098,102

\bibitem[{Potjans and Diesmann(2014)}]{PotDie14}
Potjans TC, Diesmann M (2014) The cell-type specific cortical microcircuit:
  relating structure and activity in a full-scale spiking network model. Cereb
  Cortex 24(3):785--806

\bibitem[{Pozzorini et~al(2015)Pozzorini, Mensi, Hagens, Naud, Koch, and
  Gerstner}]{PozMen15}
Pozzorini C, Mensi S, Hagens O, Naud R, Koch C, Gerstner W (2015) Automated
  high-throughput characterization of single neurons by means of simplified
  spiking models. PLoS Comput Biol 11(6):e1004,275

\bibitem[{Rosenbaum et~al(2012)Rosenbaum, Rubin, and Doiron}]{RosRub12}
Rosenbaum R, Rubin J, Doiron B (2012) Short term synaptic depression imposes a
  frequency dependent filter on synaptic information transfer. PLoS Comput Biol
  8(6)

\bibitem[{{R{\"o}ssert} et~al(2016){R{\"o}ssert}, {Pozzorini}, {Chindemi},
  {Davison}, {Eroe}, {King}, {Newton}, {Nolte}, {Ramaswamy}, {Reimann}, {Wybo},
  {Gewaltig}, {Gerstner}, {Markram}, {Segev}, and {Muller}}]{RosPoz16}
{R{\"o}ssert} C, {Pozzorini} C, {Chindemi} G, {Davison} AP, {Eroe} C, {King} J,
  {Newton} TH, {Nolte} M, {Ramaswamy} S, {Reimann} MW, {Wybo} W, {Gewaltig} MO,
  {Gerstner} W, {Markram} H, {Segev} I, {Muller} E (2016) {Automated
  point-neuron simplification of data-driven microcircuit models}. ArXiv
  e-prints \eprint{1604.00087}

\bibitem[{Rubin et~al(2015)Rubin, Van~Hooser, and Miller}]{RubVan15}
Rubin DB, Van~Hooser SD, Miller KD (2015) The stabilized supralinear network: a
  unifying circuit motif underlying multi-input integration in sensory cortex.
  Neuron 85(2):402--417

\bibitem[{Schwalger et~al(2015)Schwalger, Droste, and Lindner}]{SchDro15}
Schwalger T, Droste F, Lindner B (2015) Statistical structure of neural spiking
  under non-poissonian or other non-white stimulation. J Comput Neurosci
  39(1):29--51, \doi{10.1007/s10827-015-0560-x}

\bibitem[{{Schwalger} et~al(2017){Schwalger}, {Deger}, and
  {Gerstner}}]{SchDeg17}
{Schwalger} T, {Deger} M, {Gerstner} W (2017) {Towards a theory of cortical
  columns: From spiking neurons to interacting neural populations of finite
  size}. PLoS Comput Biol 13(4):e1005,507, \doi{10.1371/journal.pcbi.1005507}

\bibitem[{Shpiro et~al(2009)Shpiro, Moreno-Bote, Rubin, and Rinzel}]{ShpMor09}
Shpiro A, Moreno-Bote R, Rubin N, Rinzel J (2009) Balance between noise and
  adaptation in competition models of perceptual bistability. J Comput Neurosci
  27(1):37--54

\bibitem[{Tsodyks et~al(1998)Tsodyks, Pawelzik, and Markram}]{TsoPaw98}
Tsodyks M, Pawelzik K, Markram H (1998) Neural networks with dynamic synapses.
  Neural Comput 10(4):821--835

\bibitem[{Tsodyks and Markram(1997)}]{TsoMar97}
Tsodyks MV, Markram H (1997) The neural code between neocortical pyramidal
  neurons depends on neurotransmitter release probability. Proc Natl Acad Sci
  USA 94:719

\bibitem[{Wilson and Cowan(1972)}]{WilCow72}
Wilson HR, Cowan JD (1972) Excitatory and inhibitory interactions in localized
  populations of model neurons. Biophys J 12(1):1

\bibitem[{Wong and Wang(2006)}]{WonWan06}
Wong KF, Wang XJ (2006) A recurrent network mechanism of time integration in
  perceptual decisions. J Neurosci 26(4):1314--1328

\bibitem[{Zucker and Regehr(2002)}]{ZucReg02}
Zucker RS, Regehr WG (2002) Short-term synaptic plasticity. Annu Rev Physiol
  64:355--405

\end{thebibliography}
\end{document}